\pdfoutput=1

\documentclass[12pt,a4paper]{article}

\usepackage{ifthen} %
\usepackage{placeins}
\newboolean{pdflatex}
\setboolean{pdflatex}{true} %

\newboolean{articletitles}
\setboolean{articletitles}{true} %

\newboolean{uprightparticles}
\setboolean{uprightparticles}{false} %

\def\paperauthors{LHCb collaboration} %
\def\paperasciititle{Searches for 25 rare and forbidden decays of D+ and Ds+ mesons} %
\def\papertitle{Searches for 25 rare and forbidden decays of \Dp and \Ds mesons} %
\def\paperkeywords{{High Energy Physics}, {LHCb}} %
\def\papercopyright{\the\year\ CERN for the benefit of the LHCb collaboration} %
\def\paperlicence{CC BY 4.0 licence}
\def\paperlicenceurl{https://creativecommons.org/licenses/by/4.0/}

\usepackage[top=1in, bottom=1.25in, left=1in, right=1in]{geometry}

\usepackage{microtype}
\usepackage{lineno}  %
\usepackage{xspace} %
\usepackage{caption} %

\usepackage{graphicx}  %
\usepackage{color}
\usepackage{colortbl}
\graphicspath{{./figs/}} %
\DeclareGraphicsExtensions{.pdf,.PDF,png,.PNG}

\usepackage{amsmath} %
\usepackage{amssymb}
\usepackage{amsfonts}
\usepackage{upgreek} %

\newcommand*\patchAmsMathEnvironmentForLineno[1]{%
\expandafter\let\csname old#1\expandafter\endcsname\csname #1\endcsname
\expandafter\let\csname oldend#1\expandafter\endcsname\csname
end#1\endcsname
 \renewenvironment{#1}%
   {\linenomath\csname old#1\endcsname}%
   {\csname oldend#1\endcsname\endlinenomath}%
}
\newcommand*\patchBothAmsMathEnvironmentsForLineno[1]{%
  \patchAmsMathEnvironmentForLineno{#1}%
  \patchAmsMathEnvironmentForLineno{#1*}%
}
\AtBeginDocument{%
\patchBothAmsMathEnvironmentsForLineno{equation}%
\patchBothAmsMathEnvironmentsForLineno{align}%
\patchBothAmsMathEnvironmentsForLineno{flalign}%
\patchBothAmsMathEnvironmentsForLineno{alignat}%
\patchBothAmsMathEnvironmentsForLineno{gather}%
\patchBothAmsMathEnvironmentsForLineno{multline}%
\patchBothAmsMathEnvironmentsForLineno{eqnarray}%
}

\usepackage{hyperxmp}

\usepackage[pdftex,
            pdfauthor={\paperauthors},
            pdftitle={\paperasciititle},
            pdfkeywords={\paperkeywords},
            pdfcopyright={Copyright (C) \papercopyright},
            pdflicenseurl={\paperlicenceurl}]{hyperref}

\usepackage[colorinlistoftodos,textsize=scriptsize]{todonotes}

\usepackage[all]{hypcap} %

\usepackage{xspace} 
\usepackage{upgreek}

\def\lhcb   {\mbox{LHCb}\xspace}

\def\MagUp {\mbox{\em Mag\kern -0.05em Up}\xspace}

\ifthenelse{\boolean{uprightparticles}}%
{

 \def\Peta        {\ensuremath{\upeta}\xspace}

 \def\Pmu         {\ensuremath{\upmu}\xspace}

 \def\Ppi         {\ensuremath{\uppi}\xspace}                 
                  
 \def\Prho        {\ensuremath{\uprho}\xspace}

 \def\Pphi        {\ensuremath{\upphi}\xspace}

 \def\Ppsi        {\ensuremath{\uppsi}\xspace}                 
 \def\Pomega      {\ensuremath{\upomega}\xspace}                 

 \def\PDelta      {\ensuremath{\Delta}\xspace}                 
 \def\PXi         {\ensuremath{\Xi}\xspace}                 
 \def\PLambda     {\ensuremath{\Lambda}\xspace}                 
 \def\PSigma      {\ensuremath{\Sigma}\xspace}                 
 \def\POmega      {\ensuremath{\Omega}\xspace}                 
 \def\PUpsilon    {\ensuremath{\Upsilon}\xspace}

 \def\PB      {\ensuremath{\mathrm{B}}\xspace}                 
                  
 \def\PD      {\ensuremath{\mathrm{D}}\xspace}

 \def\PJ      {\ensuremath{\mathrm{J}}\xspace}                 
 \def\PK      {\ensuremath{\mathrm{K}}\xspace}

 \def\Pb      {\ensuremath{\mathrm{b}}\xspace}                 
 \def\Pc      {\ensuremath{\mathrm{c}}\xspace}                 
                  
 \def\Pe      {\ensuremath{\mathrm{e}}\xspace}

 \def\Ph      {\ensuremath{\mathrm{h}}\xspace}                 
 \def\Pi      {\ensuremath{\mathrm{i}}\xspace}

 \def\Pl      {\ensuremath{\mathrm{l}}\xspace}

 \def\Pp      {\ensuremath{\mathrm{p}}\xspace}

 \def\Ps      {\ensuremath{\mathrm{s}}\xspace}                 
                  
 \def\Pu      {\ensuremath{\mathrm{u}}\xspace}

 \def\thebaroffset{0.0em}
}
{

 \def\Peta        {\ensuremath{\eta}\xspace}

 \def\Pmu         {\ensuremath{\mu}\xspace}

 \def\Ppi         {\ensuremath{\pi}\xspace}                 
                  
 \def\Prho        {\ensuremath{\rho}\xspace}

 \def\Pphi        {\ensuremath{\phi}\xspace}

 \def\Ppsi        {\ensuremath{\psi}\xspace}                 
 \def\Pomega      {\ensuremath{\omega}\xspace}                 
 \mathchardef\PDelta="7101
 \mathchardef\PXi="7104
 \mathchardef\PLambda="7103
 \mathchardef\PSigma="7106
 \mathchardef\POmega="710A
 \mathchardef\PUpsilon="7107
                  
 \def\PB      {\ensuremath{B}\xspace}                 
                  
 \def\PD      {\ensuremath{D}\xspace}

 \def\PJ      {\ensuremath{J}\xspace}                 
 \def\PK      {\ensuremath{K}\xspace}

 \def\Pb      {\ensuremath{b}\xspace}                 
 \def\Pc      {\ensuremath{c}\xspace}                 
                  
 \def\Pe      {\ensuremath{e}\xspace}

 \def\Ph      {\ensuremath{h}\xspace}                 
 \def\Pi      {\ensuremath{i}\xspace}

 \def\Pl      {\ensuremath{l}\xspace}

 \def\Pp      {\ensuremath{p}\xspace}

 \def\Ps      {\ensuremath{s}\xspace}                 
                  
 \def\Pu      {\ensuremath{u}\xspace}

 \def\thebaroffset{0.18em}
}
\newcommand{\offsetoverline}[2][\thebaroffset]{\kern #1\overline{\kern -#1 #2}}%

\makeatletter
\ifcase \@ptsize \relax%
  \newcommand{\miniscule}{\@setfontsize\miniscule{4}{5}}%
\or%
  \newcommand{\miniscule}{\@setfontsize\miniscule{5}{6}}%
\or%
  \newcommand{\miniscule}{\@setfontsize\miniscule{5}{6}}%
\fi
\makeatother

\DeclareRobustCommand{\optbar}[1]{\shortstack{{\miniscule (\rule[.5ex]{1.25em}{.18mm})}
  \\ [-.7ex] $#1$}}

\def\en         {{\ensuremath{\Pe^-}}\xspace}   %
\def\ep         {{\ensuremath{\Pe^+}}\xspace}

\def\mup        {{\ensuremath{\Pmu^+}}\xspace}
\def\mun        {{\ensuremath{\Pmu^-}}\xspace} %

\def\ellell     {\ensuremath{\ell^+ \ell^-}\xspace}

\def\uquark    {{\ensuremath{\Pu}}\xspace}

\def\squark    {{\ensuremath{\Ps}}\xspace}

\def\cquark    {{\ensuremath{\Pc}}\xspace}

\def\bquark    {{\ensuremath{\Pb}}\xspace}

\def\hadron {{\ensuremath{\Ph}}\xspace}
\def\pion   {{\ensuremath{\Ppi}}\xspace}

\def\pip    {{\ensuremath{\pion^+}}\xspace}
\def\pim    {{\ensuremath{\pion^-}}\xspace}

\def\rhomeson {{\ensuremath{\Prho}}\xspace}
\def\rhoz     {{\ensuremath{\rhomeson^0}}\xspace}

\def\kaon    {{\ensuremath{\PK}}\xspace}

\def\KorKbar {\kern \thebaroffset\optbar{\kern -\thebaroffset \PK}{}\xspace}

\def\Kp      {{\ensuremath{\kaon^+}}\xspace}
\def\Km      {{\ensuremath{\kaon^-}}\xspace}

\def\D       {{\ensuremath{\PD}}\xspace}

\def\DorDbar {\kern \thebaroffset\optbar{\kern -\thebaroffset \PD}\xspace}

\def\Dp      {{\ensuremath{\D^+}}\xspace}

\def\Ds      {{\ensuremath{\D^+_\squark}}\xspace}
\def\Dsp     {{\ensuremath{\D^+_\squark}}\xspace}

\def\B       {{\ensuremath{\PB}}\xspace}

\def\BorBbar {\kern \thebaroffset\optbar{\kern -\thebaroffset \PB}\xspace}

\def\Bd      {{\ensuremath{\B^0}}\xspace}

\def\BdorBdbar {\kern \thebaroffset\optbar{\kern -\thebaroffset \Bd}\xspace}

\def\Bs      {{\ensuremath{\B^0_\squark}}\xspace}

\def\BsorBsbar {\kern \thebaroffset\optbar{\kern -\thebaroffset \Bs}\xspace}

\def\jpsi     {{\ensuremath{{\PJ\mskip -3mu/\mskip -2mu\Ppsi\mskip 2mu}}}\xspace}

\def\Y#1S{\ensuremath{\PUpsilon{(#1S)}}\xspace}

\def\proton      {{\ensuremath{\Pp}}\xspace}

\def\LorLbar     {\kern \thebaroffset\optbar{\kern -\thebaroffset \PLambda}\xspace}

\newcommand{\decay}[2]{\ensuremath{#1\!\to #2}\xspace}         %

\def\to                 {\ensuremath{\rightarrow}\xspace}

\def\order   {{\ensuremath{\mathcal{O}}}\xspace}

\def\AT#1     {\ensuremath{A_{\mathrm{T}}^{#1}}\xspace}           %

\def\C#1      {\ensuremath{\mathcal{C}_{#1}}\xspace}                       %
\def\Cp#1     {\ensuremath{\mathcal{C}_{#1}^{'}}\xspace}                    %
\def\Ceff#1   {\ensuremath{\mathcal{C}_{#1}^{\mathrm{(eff)}}}\xspace}        %
\def\Cpeff#1  {\ensuremath{\mathcal{C}_{#1}^{'\mathrm{(eff)}}}\xspace}       %
\def\Ope#1    {\ensuremath{\mathcal{O}_{#1}}\xspace}                       %
\def\Opep#1   {\ensuremath{\mathcal{O}_{#1}^{'}}\xspace}                    %

\newcommand{\nospaceunit}[1]{\ensuremath{\text{#1}}}       
\newcommand{\aunit}[1]{\ensuremath{\text{\,#1}}}       

\newcommand{\tev}{\aunit{Te\kern -0.1em V}\xspace}
\newcommand{\gev}{\aunit{Ge\kern -0.1em V}\xspace}
\newcommand{\mev}{\aunit{Me\kern -0.1em V}\xspace}
\newcommand{\kev}{\aunit{ke\kern -0.1em V}\xspace}
\newcommand{\ev}{\aunit{e\kern -0.1em V}\xspace}
\newcommand{\mevc}{\ensuremath{\aunit{Me\kern -0.1em V\!/}c}\xspace}
\newcommand{\gevc}{\ensuremath{\aunit{Ge\kern -0.1em V\!/}c}\xspace}
\newcommand{\mevcc}{\ensuremath{\aunit{Me\kern -0.1em V\!/}c^2}\xspace}
\newcommand{\gevcc}{\ensuremath{\aunit{Ge\kern -0.1em V\!/}c^2}\xspace}

\def\mum  {\ensuremath{\,\upmu\nospaceunit{m}}\xspace}

\def\barn{\aunit{b}\xspace}

\def\fb   {\ensuremath{\aunit{fb}}\xspace}
\def\invfb   {\ensuremath{\fb^{-1}}\xspace}

\def\order{{\ensuremath{\mathcal{O}}}\xspace}

\def\gsim{{~\raise.15em\hbox{$>$}\kern-.85em
          \lower.35em\hbox{$\sim$}~}\xspace}
\def\lsim{{~\raise.15em\hbox{$<$}\kern-.85em
          \lower.35em\hbox{$\sim$}~}\xspace}

\def\sPlot{\mbox{\em sPlot}\xspace}

\def\pt         {\ensuremath{p_{\mathrm{T}}}\xspace}

\def\ptot       {\ensuremath{p}\xspace}

\def\evtgen     {\mbox{\textsc{EvtGen}}\xspace}

\def\geant      {\mbox{\textsc{Geant4}}\xspace}

\def\photos     {\mbox{\textsc{Photos}}\xspace}

\def\pythia     {\mbox{\textsc{Pythia}}\xspace}

\xspace

\def\tell1  {TELL1\xspace}
\def\ukl1   {UKL1\xspace}

\def\DorDsp  {{\ensuremath{\D^+_{(s)}}}\xspace}
\def\DporDsp  {{\ensuremath{\D^+_{(s)}}}\xspace}

\def\cToull{\decay{\cquark}{\uquark \ell^{+}\ell^{\prime-}}}

\def\DTohll     {\decay{\DorDsp}{\Ph^{\pm}\ell^{+}\ell^{(\prime)\mp}}}

\def\DToKllOS   {\decay{\DorDsp}{\Kp\ell^{+}\ell^{(\prime)-}}}

\def\DTopill    {\decay{\DorDsp}{\pip\ell^{\pm}\ell^{(\prime)\mp}}}

\def\DOnlyTohll     {\decay{\Dp}{\Ph^{\pm}\ell^{+}\ell^{(\prime)\mp}}}

\def\DsOnlyTohll     {\decay{\Dsp}{\Ph^{\pm}\ell^{+}\ell^{(\prime)\mp}}}

\def\DpTopimumuOS   {\decay{\Dp}{\pip\mup\mun}}

\def\DpTopieeOS   {\decay{\Dp}{\pip\ep\en}}

\def\DspToKmumuSS   {\decay{\Dsp}{\Km\mup\mup}}
\def\DspToKmumuOS   {\decay{\Dsp}{\Kp\mup\mun}}
\def\DspToKmueSS   {\decay{\Dsp}{\Km\mup\ep}}

\def\DspToKeeSS   {\decay{\Dsp}{\Km\ep\ep}}
\def\DspToKeeOS   {\decay{\Dsp}{\Kp\ep\en}}

\def\DToKmumuSS   {\decay{\DorDsp}{\Km\mup\mup}}
\def\DToKmumuOS   {\decay{\DorDsp}{\Kp\mup\mun}}
\def\DToKmueSS   {\decay{\DorDsp}{\Km\mup\ep}}
\def\DToKmueOS   {\decay{\DorDsp}{\Kp\mup\en}}
\def\DToKemuOS   {\decay{\DorDsp}{\Kp\ep\mun}}
\def\DToKeeSS   {\decay{\DorDsp}{\Km\ep\ep}}
\def\DToKeeOS   {\decay{\DorDsp}{\Kp\ep\en}}

\def\DTopimumuSS   {\decay{\DorDsp}{\pim\mup\mup}}
\def\DTopimumuOS   {\decay{\DorDsp}{\pip\mup\mun}}
\def\DTopimueSS   {\decay{\DorDsp}{\pim\mup\ep}}
\def\DTopimueOS   {\decay{\DorDsp}{\pip\mup\en}}
\def\DTopiemuOS   {\decay{\DorDsp}{\pip\ep\mun}}
\def\DTopieeSS   {\decay{\DorDsp}{\pim\ep\ep}}
\def\DTopieeOS   {\decay{\DorDsp}{\pip\ep\en}}

\def\DToKpipiOS    {\decay{\DorDsp}{\Kp\pip\pim}}

\def\DTopipipi     {\decay{\DorDsp}{\pip\pip\pim}}

\def\DTopiphiphiToKK   {\decay{\DorDsp}{\left(\decay{\phi}{\Kp\Km}\right)} \pip}

\def\DTopiphiphiToll {\decay{\DorDsp}{\left(\decay{\phi}{\ell^-\ell^+}\right)} \pip}

\def\DTopiphiphiTomumu   {\decay{\DorDsp}{\left(\decay{\phi}{\mun\mup}\right)} \pip}
\def\DpTopiphiphiTomumu   {\decay{\Dp}{\left(\decay{\phi}{\mun\mup}\right)} \pip}
\def\DspTopiphiphiTomumu   {\decay{\Dsp}{\left(\decay{\phi}{\mun\mup}\right)} \pip}

\def\DTopiphiphiToee   {\decay{\DorDsp}{\left(\decay{\phi}{\en\ep}\right)} \pip}
\def\DpTopiphiphiToee   {\decay{\Dp}{\left(\decay{\phi}{\en\ep}\right)} \pip}
\def\DspTopiphiphiToee   {\decay{\Dsp}{\left(\decay{\phi}{\en\ep}\right)} \pip}

\def\phiToee   {\decay{\phi}{\en\ep}}
\def\phiTomumu   {\decay{\phi}{\mun\mup}}
\def\phiToKK   {\decay{\phi}{\Kp\Km}}
\def\phiToll   {\decay{\phi}{\ell^+ \ell^-}}

\def\JPsiTomumu  {\decay{\jpsi}{\mup\mun}}

\usepackage{template/cite} %
\usepackage{template/mciteplus}

\usepackage{siunitx}
\usepackage{booktabs}
\usepackage{multirow}
\usepackage{subcaption}
\usepackage{rotating}

\begin{document}

\renewcommand{\thefootnote}{\fnsymbol{footnote}}
\setcounter{footnote}{1}

\begin{titlepage}
\pagenumbering{roman}

\vspace*{-1.5cm}
\centerline{\large EUROPEAN ORGANIZATION FOR NUCLEAR RESEARCH (CERN)}
\vspace*{1.5cm}
\noindent
\begin{tabular*}{\linewidth}{lc@{\extracolsep{\fill}}r@{\extracolsep{0pt}}}
\ifthenelse{\boolean{pdflatex}}%
{\vspace*{-1.5cm}\mbox{\!\!\!\includegraphics[width=.14\textwidth]{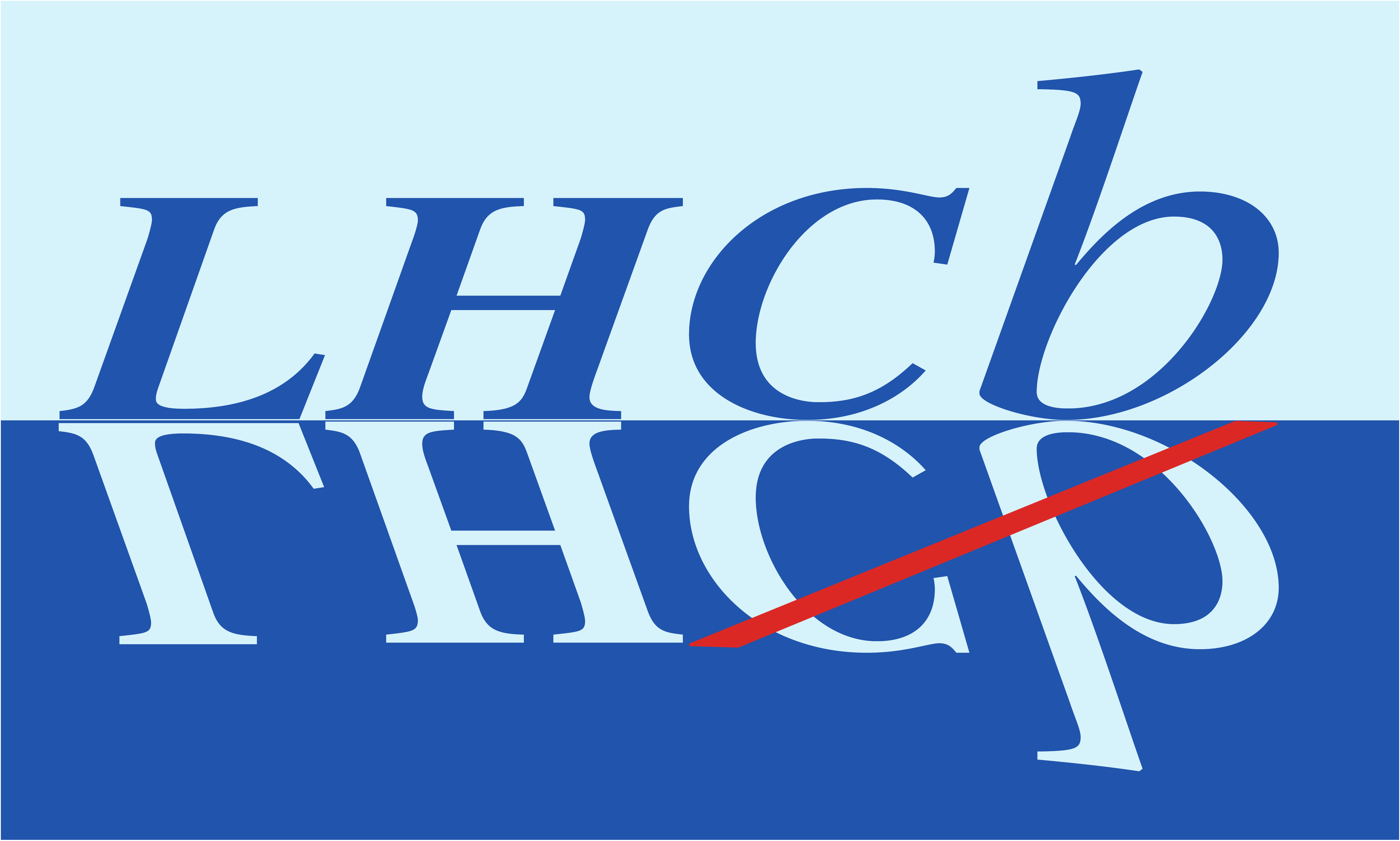}} & &}%
{\vspace*{-1.2cm}\mbox{\!\!\!\includegraphics[width=.12\textwidth]{figs/lhcb-logo.eps}} & &}%
\\
 & & CERN-EP-2020-140 \\  %
 & & LHCb-PAPER-2020-007 \\  %
 & & October 31, 2020 \\ %
 & & \\
\end{tabular*}

\vspace*{4.0cm}

{\normalfont\bfseries\boldmath\huge
\begin{center}
  \papertitle 
\end{center}
}

\vspace*{2.0cm}

\begin{center}
\paperauthors\footnote{Authors are listed at the end of this paper.}
\end{center}

\vspace{\fill}

\begin{abstract}
  \noindent
A search is performed for rare and forbidden charm decays of the form \mbox{\DTohll}, where $\smash{\hadron^\pm}$ is a pion or kaon and $\smash{\ell^{(')\pm}}$ is an electron or muon.
The measurements are performed using proton-proton collision data, corresponding to an integrated luminosity of \SI{1.6}{\per\femto\barn}, collected by the \lhcb experiment in 2016.
No evidence is observed for the 25 decay modes that are investigated and \SI{90}{\percent} confidence level limits on the branching fractions are set between $1.4\times10^{-8}$ and $6.4\times10^{-6}$.
In most cases, these results represent an improvement on existing limits by one to two orders of magnitude.

\end{abstract}

\vspace*{2.0cm}

\begin{center}
  Submitted to JHEP
\end{center}

\vspace{\fill}

{\footnotesize 
\centerline{\copyright~\papercopyright. \href{\paperlicenceurl}{\paperlicence}.}}
\vspace*{2mm}

\end{titlepage}

\newpage
\setcounter{page}{2}
\mbox{~}

\cleardoublepage

\renewcommand{\thefootnote}{\arabic{footnote}}
\setcounter{footnote}{0}

\pagestyle{plain} %
\setcounter{page}{1}
\pagenumbering{arabic}

\section{Introduction}
\label{sec:Introduction}

The search for rare and forbidden decays in flavour physics constitutes an important test of the Standard Model (SM) and provides a window to new physics. In the charm sector, decays of the form \DOnlyTohll and \DsOnlyTohll, where $\Ph^\pm$ is a charged pion or kaon and $\Pl^{\left(\prime\right)\pm}$ is
an electron or muon, are among such processes.\footnote{Throughout this article the inclusion of charge-conjugate processes and the use of natural units ($c=\hbar=1$) is implied, unless otherwise indicated.} Searches are reported for 25 decay modes, with common experimental approaches applied to all channels.

The physics processes of these channels vary.
Four of the decay channels (\mbox{\DpTopieeOS}, \mbox{\DpTopimumuOS}, \mbox{\DspToKeeOS}, \mbox{\DspToKmumuOS}) involve flavour-changing neutral current (FCNC) transitions. These processes are rare within the Standard Model as only weak annihilation processes can occur at tree level.
At the loop level, FCNC transitions are suppressed by the Glashow–-Iliopoulos-–Maiani (GIM) mechanism~\cite{Glashow:1970gm} but are well established in both $B$-meson and $K$-meson decays \cite{PDG2018}.
FCNC $B$-meson decays have received significant attention in recent years: the process \decay{\Bs}{\mup\mun} was observed\cite{LHCb-PAPER-2014-049,LHCb-PAPER-2017-001}; and experimental observations in analyses of \decay{\bquark}{\squark\ellell} transitions are in tension with the SM predictions (see Ref.~\cite{LHCb-PAPER-2019-009} and references therein).
The GIM cancellation in the $D$-meson system is stronger than in the $B$-meson system, leading to short-distance SM branching fractions of $\order(10^{-12})$~\cite{deBoer:2015boa}.
Significant long-distance contributions occur from dilepton resonances and these are expected to dominate the SM contribution across the full dilepton invariant-mass-squared distribution~\cite{deBoer:2015boa, Fajfer:2015mia}.

The eight decay modes with two oppositely charged leptons of different flavour (\mbox{\DTopiemuOS}, \mbox{\DTopimueOS}, \mbox{\DToKemuOS}, \mbox{\DToKmueOS}) are lepton-flavour violating (LFV) and lepton-number conserving decays.
These decays have only negligible contributions from neutrino mixing in the SM.
The nine decay modes analysed with same-sign leptons (\mbox{\DTopieeSS},
\mbox{\DTopimumuSS},
\mbox{\DspToKeeSS},
\mbox{\DspToKmumuSS},
\mbox{\DTopimueSS},
\mbox{\DspToKmueSS}) are both LFV and lepton-number violating (LNV) decays, and forbidden in the SM.
There are three further LFV and LNV decays  (\mbox{\DspToKeeSS}, \mbox{\DspToKmumuSS}, \mbox{\DspToKmueSS}) which are analysed here, their \Dp counterparts are not studied due to the presence of sizeable background contributions.

The potential of \cToull processes to constrain new physics has been discussed in the literature~\cite{deBoer:2015boa, Fajfer:2015mia, Bause:2019vpr, Bause:2020obd}.
Model-independent constraints on Wilson coefficients in effective field theory have been considered, as have specific classes of models.
Recently particular attention has been paid to leptoquark models, due to their potential relevance to the anomalies on \decay{\bquark}{\squark\ellell} processes~\cite{Paul:2011ar,Burdman:2001tf,Wang:2014dba,Delaunay:2012cz,Paul:2012ab}.

The analysis reported here is performed on a data sample corresponding to an integrated luminosity of 1.6\invfb collected by the LHCb experiment in 2016.
Four resonant decays (\mbox{\DTopiphiphiTomumu}, \mbox{\DTopiphiphiToee}) are used for calibration and normalisation.
Non-resonant decays with a pion and two same-sign or opposite-sign muons in the final state have previously been searched for by LHCb with no evidence observed~\cite{LHCB-PAPER-2012-051}.
The best limits in the other channels are measured by the BaBar collaboration~\cite{Lees:2011hb}, which has studied all channels, or by the CLEO collaboration, from searches in decay modes with dielectrons in the final state~\cite{Rubin:2010cq}. 
\section{Detector and simulation}
\label{sec:Detector}

The \lhcb detector~\cite{LHCb-DP-2008-001,LHCb-DP-2014-002} is a single-arm forward
spectrometer covering the \mbox{pseudorapidity} range $2<\eta <5$,
designed for the study of particles containing \bquark or \cquark
quarks. The detector includes a high-precision tracking system
consisting of a silicon-strip vertex detector surrounding the $pp$
interaction region, a large-area silicon-strip detector located
upstream of a dipole magnet with a bending power of about
$4{\mathrm{\,Tm}}$, and three stations of silicon-strip detectors and straw
drift tubes placed downstream of the magnet.
The tracking system provides a measurement of the momentum, \ptot, of charged particles with
a relative uncertainty that varies from 0.5\% at low momentum to 1.0\% at 200\gev.
The minimum distance of a track to a primary \proton\proton collision vertex (PV), the impact parameter (IP), 
is measured with a resolution of $(15+29/\pt)\mum$,
where \pt is the component of the momentum transverse to the beam, in\,\gev.
Different types of charged hadrons are distinguished using information
from two ring-imaging Cherenkov detectors. 
Photons, electrons and hadrons are identified by a calorimeter system consisting of
scintillating-pad and preshower detectors, an electromagnetic
and a hadronic calorimeter. Muons are identified by a
system composed of alternating layers of iron and multiwire
proportional chambers.

Simulation is required to model the effects of the detector acceptance and the
  selection requirements.
  In the simulation, $pp$ collisions are generated using
\pythia~\cite{Sjostrand:2007gs,*Sjostrand:2006za} 
 with a specific \lhcb
configuration~\cite{LHCb-PROC-2010-056}.  Decays of unstable particles
are described by \evtgen~\cite{Lange:2001uf}, in which final-state
radiation is generated using \photos~\cite{Golonka:2005pn}. The
interaction of the generated particles with the detector, and its response,
are implemented using the \geant
toolkit~\cite{Allison:2006ve, *Agostinelli:2002hh} as described in
Ref.~\cite{LHCb-PROC-2011-006}.

\section{Triggering, reconstruction and selection}
\label{sec:Selection}

The online event selection is performed by a trigger, 
which consists of a hardware stage, based on information from the calorimeter and muon
systems, followed by a two-level software stage. In between the two software stages, an alignment and calibration of the detector is performed in near real time and the results are used in the trigger.
The same alignment and calibration information is propagated 
to the offline reconstruction, ensuring consistent and high-quality 
particle identification (PID) performance between the trigger and 
offline software.

A hardware trigger selection is made for all decay channels based on observing a deposit with high transverse energy in the hadronic calorimeter. Depending on the leptons in the final state, this is supplemented by a selection of events with high transverse-energy clusters in the electromagnetic calorimeter and high transverse-momentum muons.
At the first software level, inclusive multivariate selections are used to select charged tracks which do not originate from any PV.

The second software level builds \DporDsp candidates using exclusive selections for each combination of final-state particles.
Only events containing a single reconstructed PV are used.
All three tracks are required to be incompatible with coming from the PV, and to have been reconstructed with good fit quality, have a transverse momentum exceeding $300{\mev}$ and a momentum exceeding $2000{\mev}$.
The tracks are required to form a \DporDsp secondary vertex, with good fit quality and all pairs of tracks intercepting within \SI{150}{\micro\meter}.
The angle between the reconstructed \DporDsp and the vector connecting the PV to the decay vertex of the \DporDsp candidate (direction angle) must be within \SI{14}{\milli\radian}.
Loose PID requirements are made on all final-state tracks.
Eight of the signal channels in this analysis can proceed via decays through intermediate resonances (\Peta, \rhoz, \Pomega and \Pphi).
These are removed by vetoing the region $[\SI{525}{\mega\electronvolt},\SI{1250}{\mega\electronvolt}]$ in dilepton mass.
The resonant channel \mbox{\DTopiphiphiTomumu} (\mbox{\DTopiphiphiToee}) is further selected by requiring the invariant mass of the dilepton pair be within $\pm\num{20}$ ($^{+\num{40}}_{-\num{100}}$)~\mev of the $\phi$ mass\cite{PDG2018}.
A bremsstrahlung reconstruction procedure is used to correct the momentum of electron candidates.

A classifier trained using XGBoost\cite{xgboost}, is used to further distinguish between signal and background originating from random combinations of tracks. A single classifier is trained for each final state, with signal being represented by an equally weighted mixture of simulated \Dp and \Dsp events.
This approach is found to give equivalent performance when compared to using a separate classifier for each signal meson.
Background candidates for the training are provided from a data sample where all three final-state particles have the same electric charge.
The classifier is trained using the fit quality of the primary and secondary vertex, the \DorDsp pseudorapidity, the \DorDsp flight distance, the difference in $\chi^2$ of the PV reconstructed with and without each final-state track, the momentum of each final-state track, the maximum distance of closest approach between all final-state tracks, the direction angle and the reconstructed proper lifetime of the \DorDsp.
To guarantee that no other tracks can be associated to the secondary vertex, an isolation variable, $A_{\pt}$, is used that considers the imbalance of \pt of nearby tracks compared to that of the \DporDsp candidate,
\begin{equation}
A_{\pt} = \frac{\pt(\DporDsp)-(\sum \overrightarrow{p})_{\rm T}}{\pt(\DporDsp)+(\sum \overrightarrow{p})_{\rm T}},
\end{equation}
where $\pt(\DorDsp)$ is the \pt of the \DporDsp meson and $(\sum \overrightarrow{p})_{\rm T}$ is the transverse component of the vector sum of all charged particles momenta within a cone around the candidate, excluding the three signal tracks. The cone is defined by a circle of radius 2.0 in the plane of pseudorapidity and azimuthal angle, measured in radians, around the \DorDsp candidate direction.
The signal \DporDsp candidates tend to be more isolated than the combinatorial background and show on average greater values of $A_{\pt} $.
The final selection was optimised using a grid search in the thresholds applied to the classifier output and the PID variables.

\section{Invariant-mass distributions}
\label{massDistributions}

\begin{figure}[t]
    \centering
    \begin{subfigure}[t]{0.49\textwidth}
        \includegraphics[width=\textwidth]{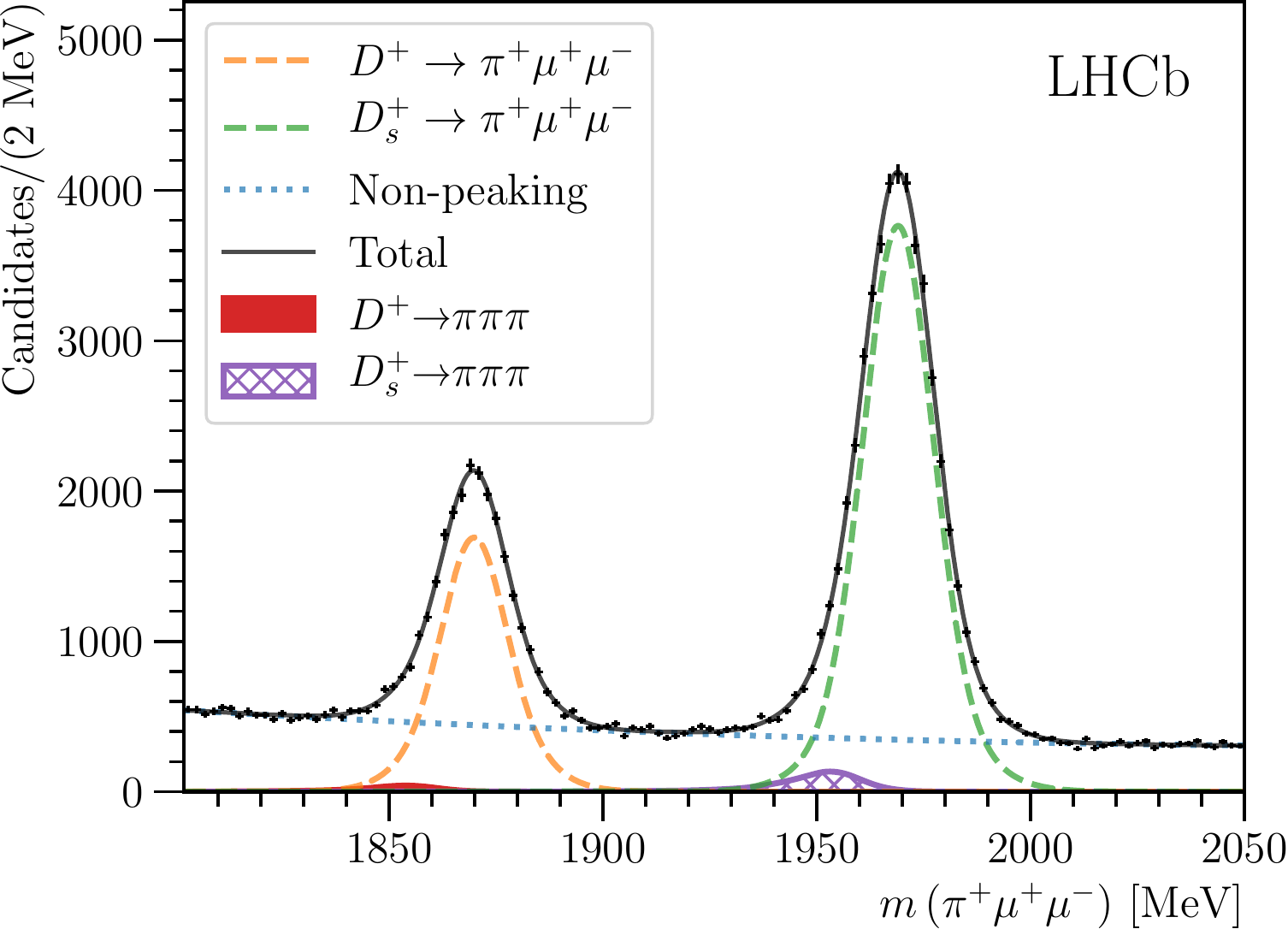}
    \end{subfigure}
    \begin{subfigure}[t]{0.49\textwidth}
        \includegraphics[width=\textwidth]{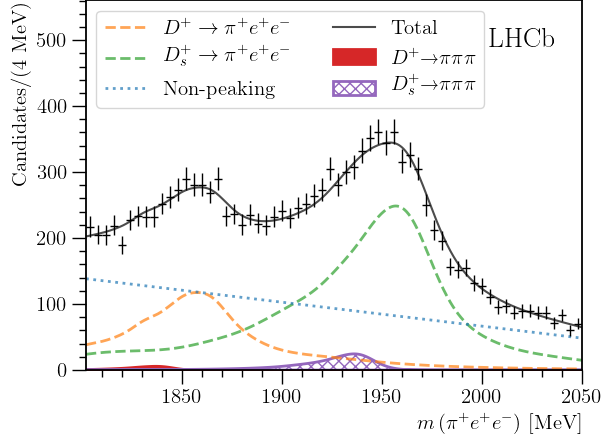}
    \end{subfigure}

    \caption{\small Distributions of the three-body invariant mass for the \DTopiphiphiToll candidates used for normalisation and calibration. The final states are specified in the mass label.
    The \Dp (\Dsp) signal component is shown as a dashed orange (green) line, peaking backgrounds are denoted by the solid (hashed) regions and the non-peaking background is denoted by the blue dotted line.}
    \label{fig:norm-fits}
\end{figure}

\begin{figure}[htp]
    \centering
    \begin{subfigure}[t]{0.49\textwidth}
        \includegraphics[width=\textwidth]{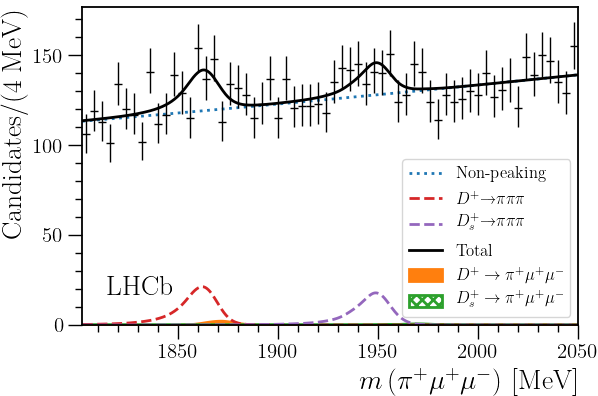}
    \end{subfigure}
    \begin{subfigure}[t]{0.49\textwidth}
        \includegraphics[width=\textwidth]{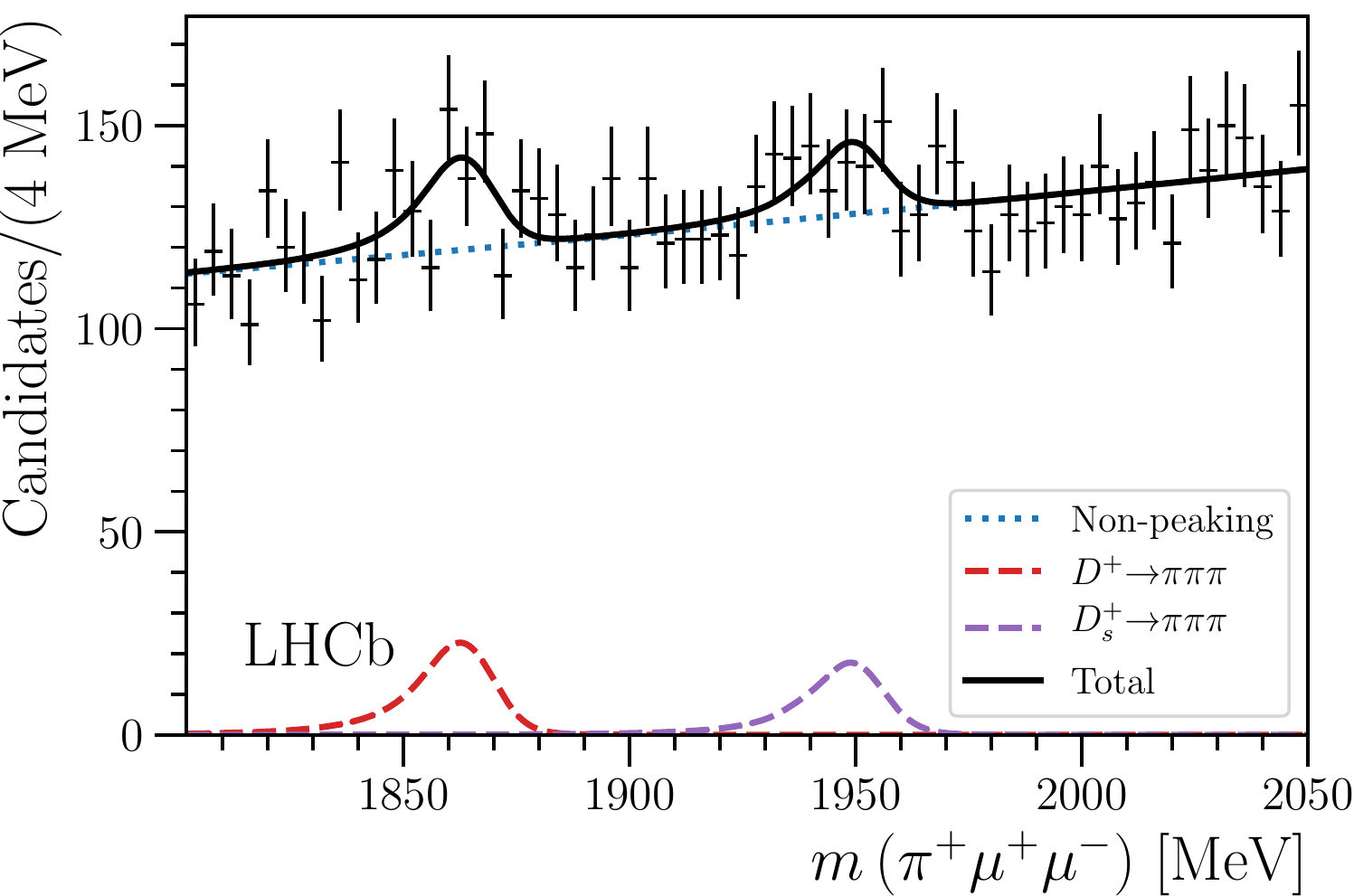}
    \end{subfigure}
    \begin{subfigure}[t]{0.49\textwidth}
        \includegraphics[width=\textwidth]{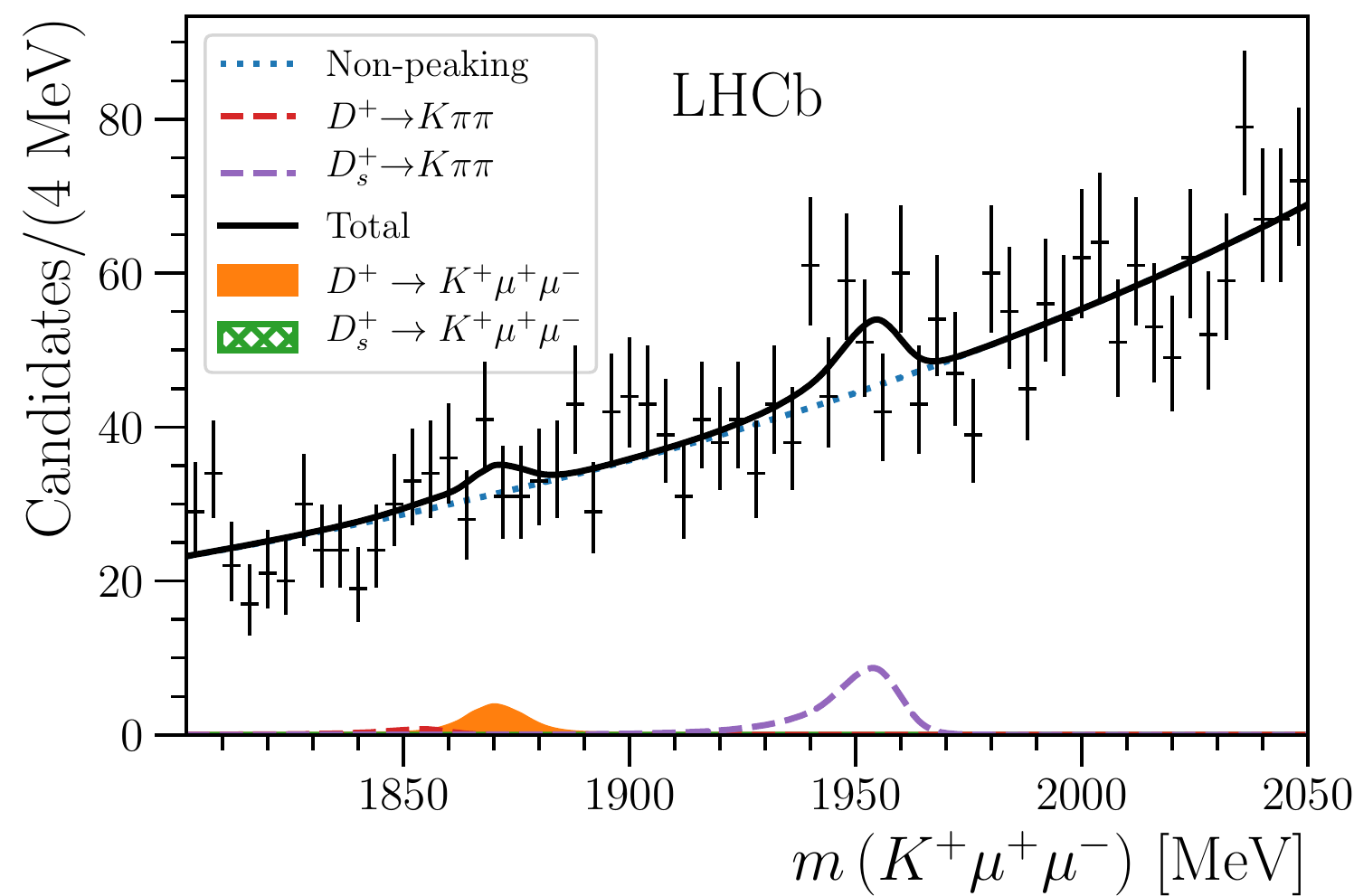}
    \end{subfigure}
    \begin{subfigure}[t]{0.49\textwidth}
        \includegraphics[width=\textwidth]{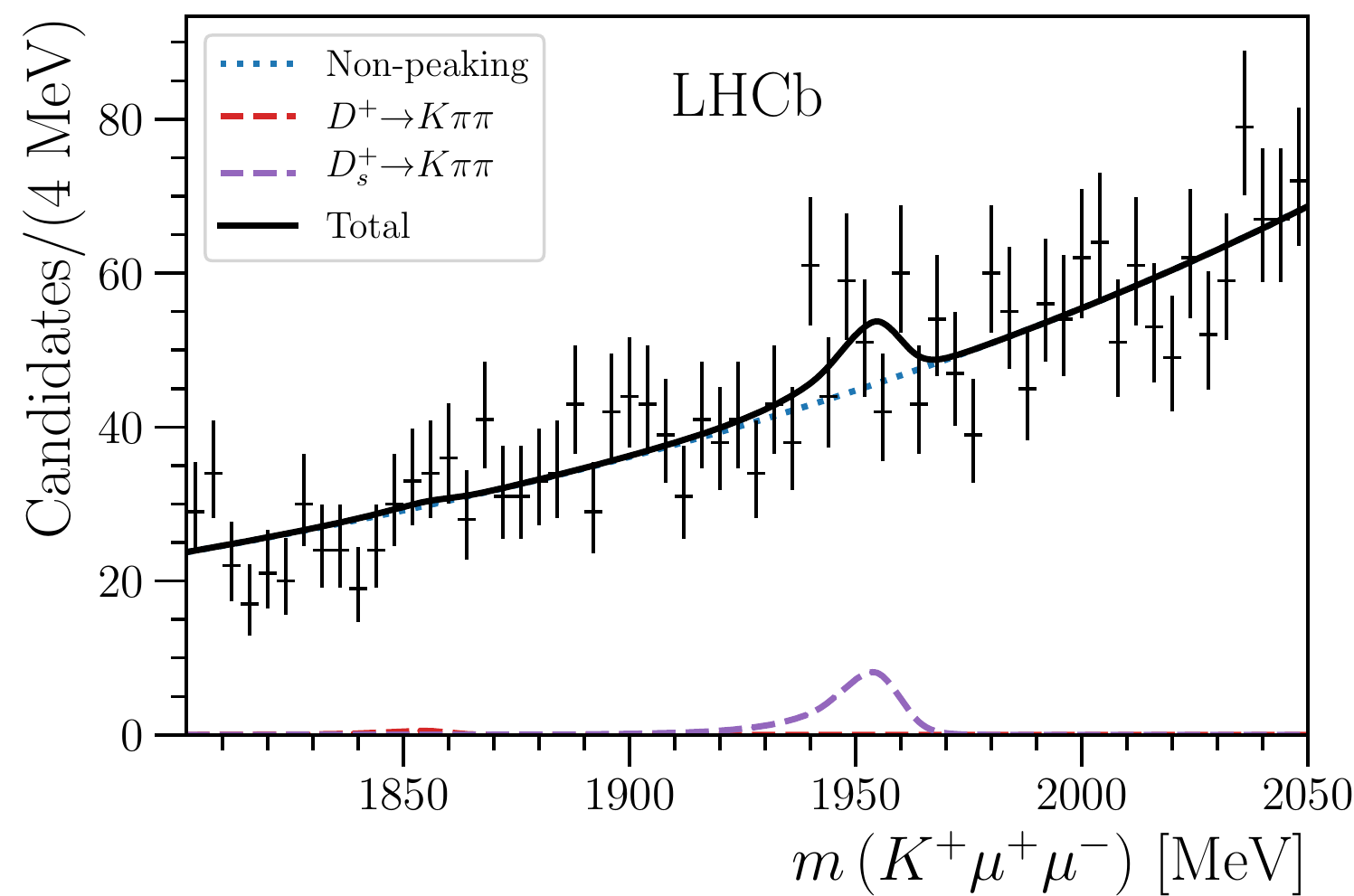}
    \end{subfigure}
    \begin{subfigure}[t]{0.49\textwidth}
        \includegraphics[width=\textwidth]{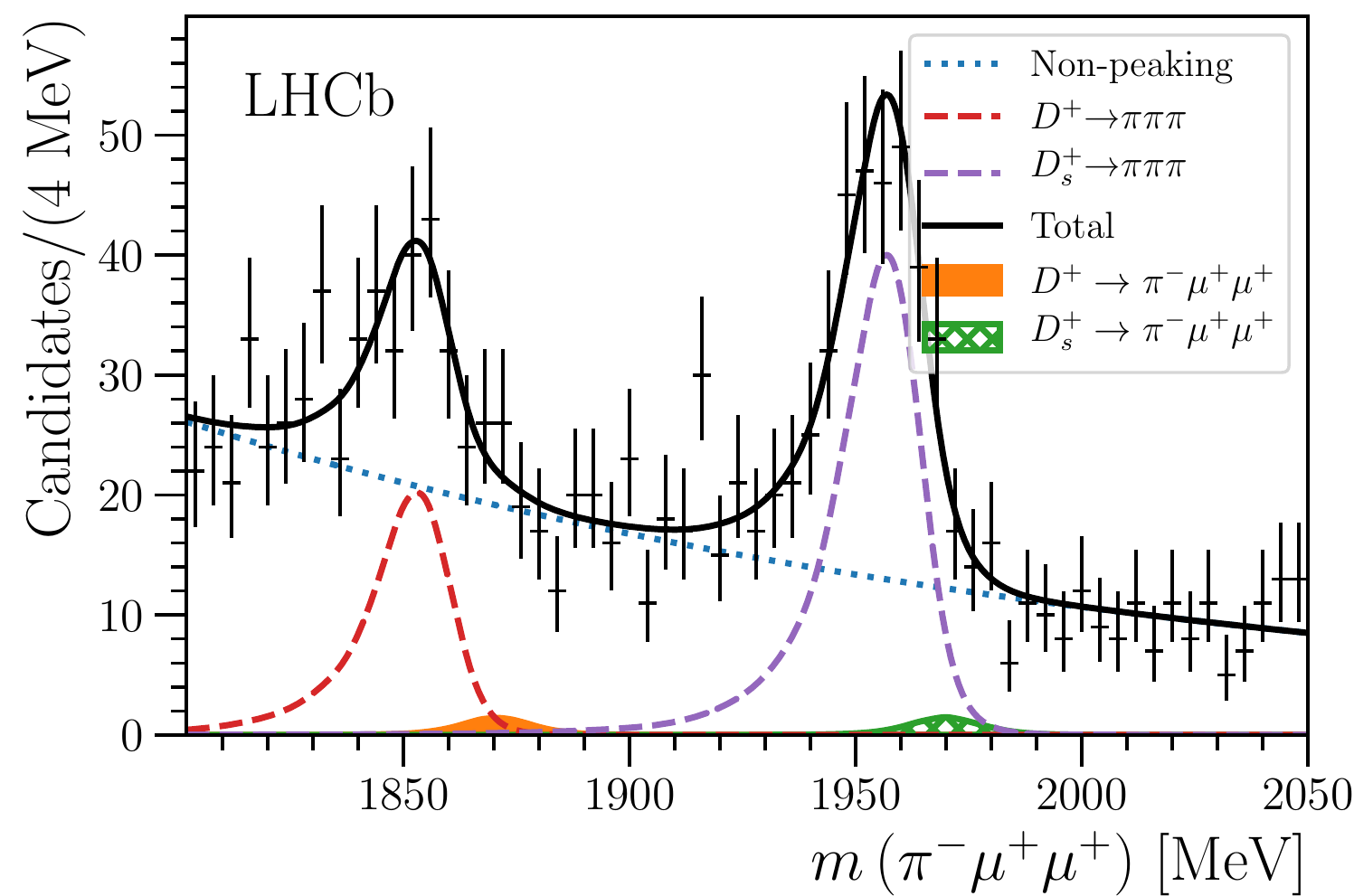}
    \end{subfigure}
    \begin{subfigure}[t]{0.49\textwidth}
        \includegraphics[width=\textwidth]{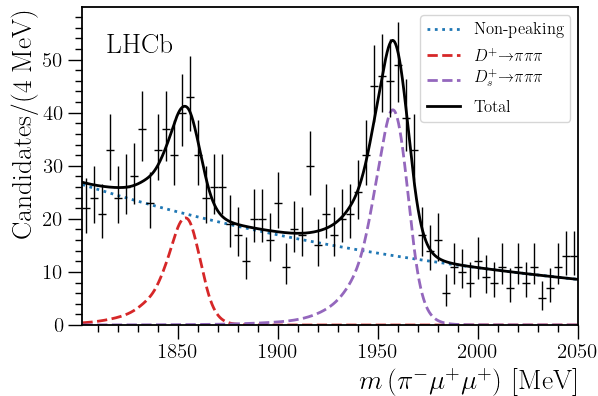}
    \end{subfigure}
    \begin{subfigure}[t]{0.49\textwidth}
        \includegraphics[width=\textwidth]{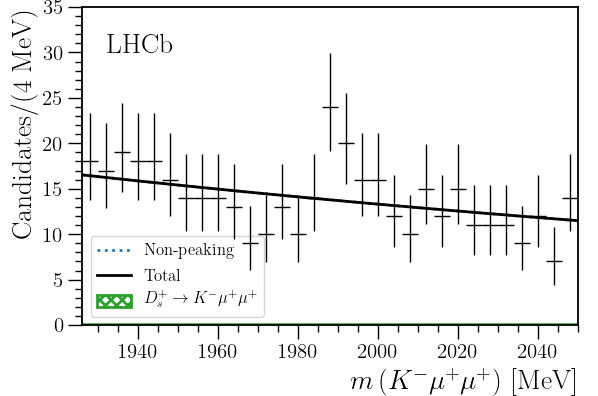}
    \end{subfigure}
    \begin{subfigure}[t]{0.49\textwidth}
        \includegraphics[width=\textwidth]{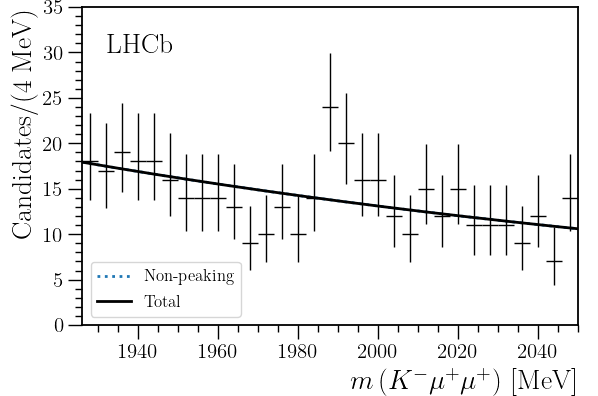}
    \end{subfigure}

    \caption{\small Distributions of the three-body invariant mass in the signal regions for decays with two muons. The final states are specified in the mass label.
    The left (right) fit with the signal-plus-background (background-only) hypothesis is overlaid with the peaking backgrounds denoted by dashed lines and the non-peaking background is denoted by the blue dotted line.
    }
    \label{fig:mass-fits:1}
\end{figure}

\begin{figure}[htp]
    \centering
    \begin{subfigure}[t]{0.49\textwidth}
        \includegraphics[width=\textwidth]{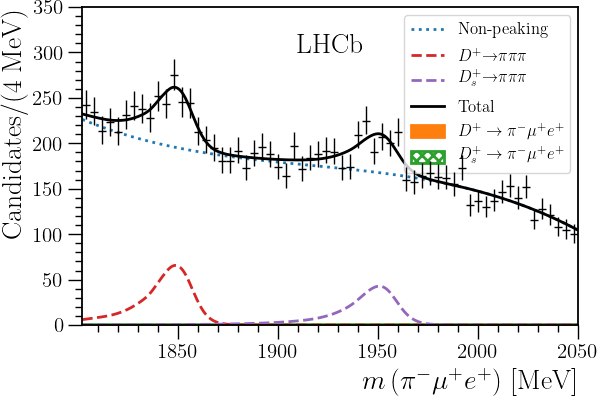}
    \end{subfigure}
    \begin{subfigure}[t]{0.49\textwidth}
        \includegraphics[width=\textwidth]{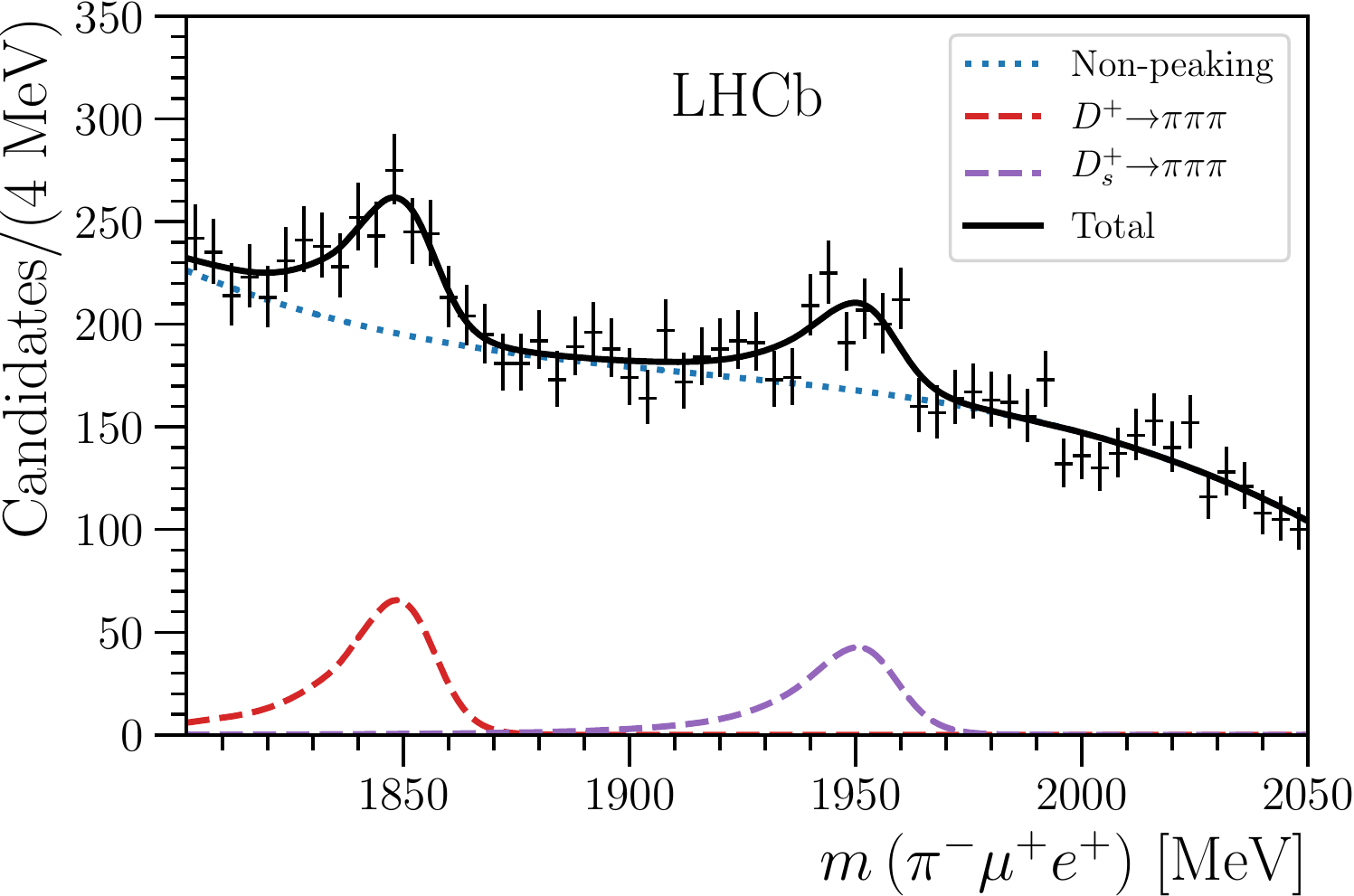}
    \end{subfigure}
    \begin{subfigure}[t]{0.49\textwidth}
        \includegraphics[width=\textwidth]{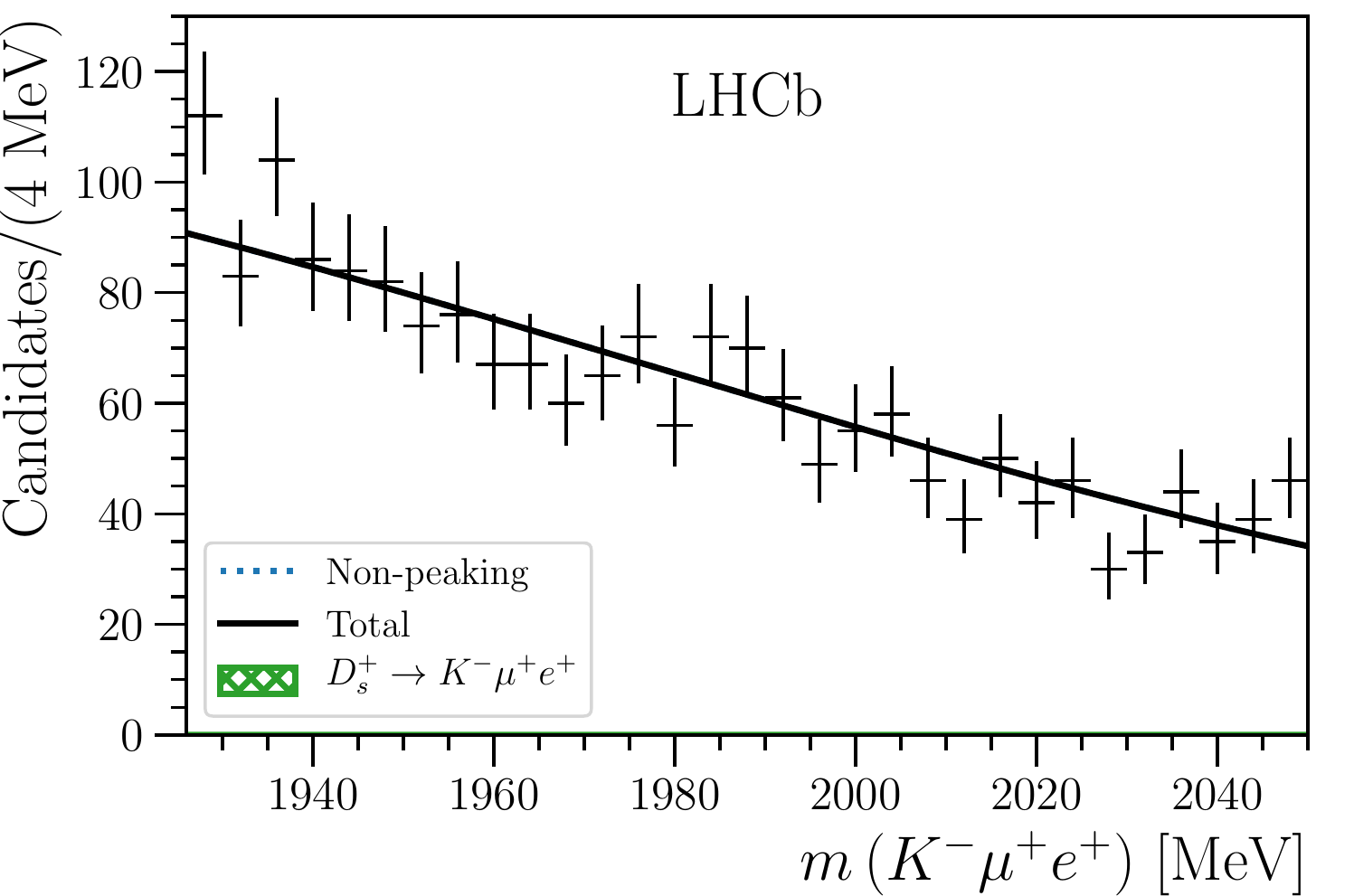}
    \end{subfigure}
    \begin{subfigure}[t]{0.49\textwidth}
        \includegraphics[width=\textwidth]{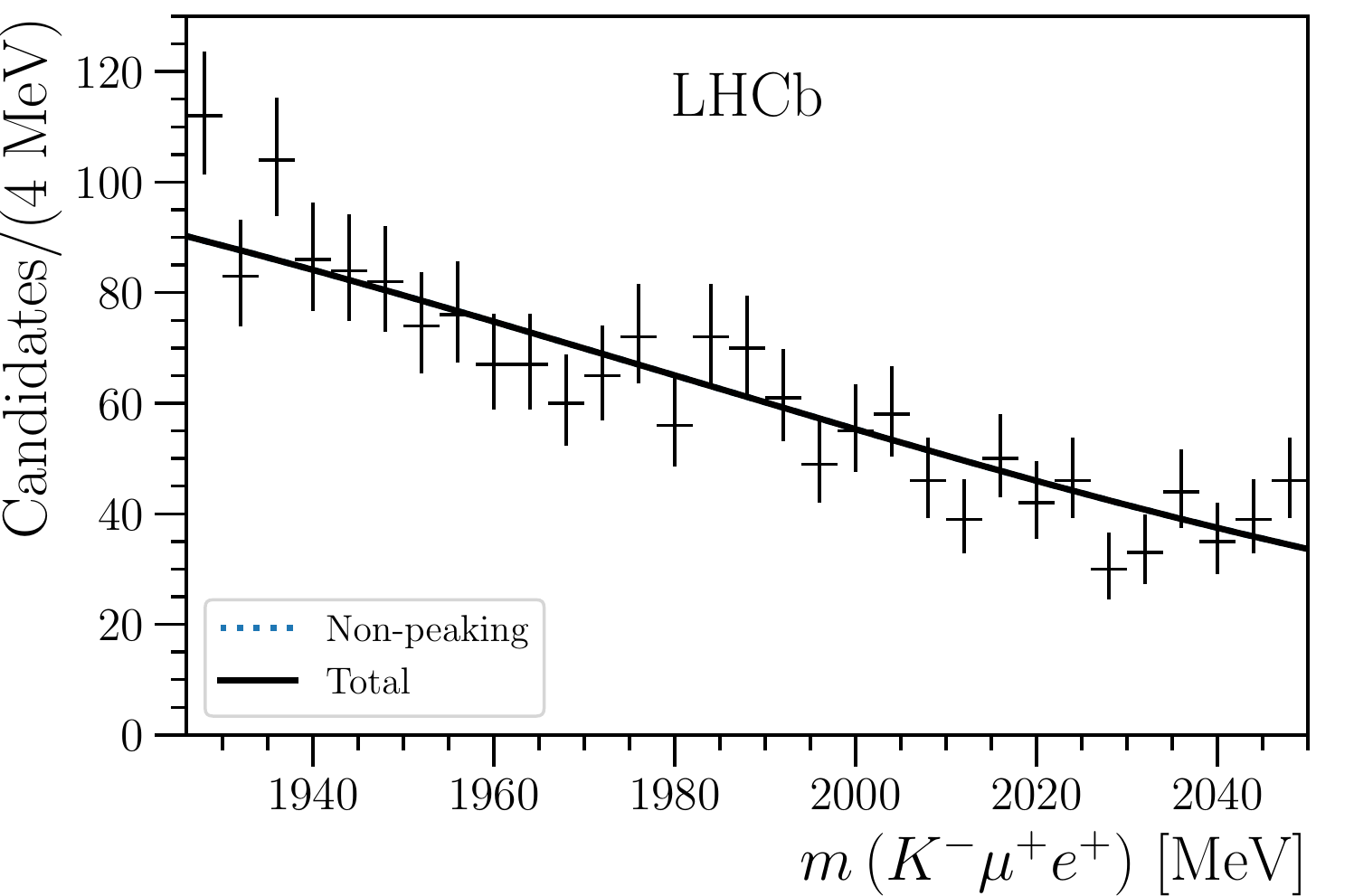}
    \end{subfigure}

    \caption{\small Distributions of the three-body invariant mass in the signal regions for decays with an electron and a muon with matching charge. The final states are specified in the mass label.
    The left (right) fit with the signal-plus-background (background-only) hypothesis is overlaid with the peaking backgrounds denoted by dashed lines and the non-peaking background is denoted by the blue dotted line.
    }
    \label{fig:mass-fits:4}
\end{figure}

The signal yields are determined using maximum-likelihood fits to the invariant-mass distributions in each of the \DTohll final states, including the resonant channels.
The fit is performed in the mass range of \num{1802} to \SI{2050}{\MeV} for all final states where both the \Dp and \Dsp decay channels are analysed.
In the final states where only the \Dsp contribution is analysed, the fit is performed in the range of \num{1926} to \SI{2050}{\MeV}.
The invariant-mass distributions are shown in Figs.~\ref{fig:norm-fits}-\ref{fig:mass-fits:4}, with fits indicated. No evidence is observed in any of the 25 signal channels.
The yields obtained from the fit to the normalisation channels can be found in Table~\ref{tab:norm-fit-result}.

The mass distributions for the signal contributions are obtained by modelling the distributions in simulated events with a kernel density estimation (KDE) technique~\cite{Cranmer:2000du}.  The same technique is applied to the \DTopiphiphiToee channels. In the  \DTopiphiphiTomumu channels, where a clean separation from background can be made, a fit with a double Gaussian distribution is performed to the data, with the two Gaussian distributions sharing a common mean.

In decay channels with electrons, the observed mass distribution is dependent on the number of reconstructed bremsstrahlung photons. For final states with one electron, separate shapes are extracted from simulation for candidates with no bremsstrahlung and where one or more reconstructed photon candidates are associated with the electron. For final states with two electrons, separate shapes are used for candidates with no bremsstrahlung, one photon per \DporDsp candidate, and two or more photons per \DporDsp candidate.

Background arises from random combinations which vary smoothly across the fit range, and from misidentified and partially reconstructed backgrounds that may peak in the fitted mass range. The combinatorial backgrounds are fitted with an exponential distribution where the analysed signal channel contains dimuons, while third-order Chebyshev polynomials are used in the cases where the signal channel contains one or two electrons. The form used for modelling these combinatorial backgrounds is determined using a data sample where all three final-state tracks have the same charge.

Background from decays involving leptons and neutrinos are found to be negligible for this analysis, with only three-body \Dp and \Dsp hadronic decays  contributing to the signal samples.
The backgrounds that affect a specific signal channel depend on whether the signal-channel hadron is a pion or a kaon and whether the leptons are of the same or opposite charge.
Simulation samples, which do not use a description of the interaction of the particles with the detector material, are generated~\cite{Cowan:2016tnm}, and
are fitted using the same KDE technique as the signal.

The modelling of the backgrounds is validated using data samples where the nominal kinematic selection is applied, but with the particle identification requirements reversed and their values modified from those in the signal selection to enhance the background.
The distributions in these enhanced background samples are then fitted with the shapes obtained from the KDE fit to the simplified simulation plus a combinatorial-background description.
The four final states of the form \DToKllOS have significant contributions from \DToKpipiOS decays, and
the seven final states of the form \DTopill have significant contributions from \DTopipipi decays.
In these cases, the KDE description is smeared by a Gaussian function with mean and width included as free parameters of the fit.
After applying the additional smearing a good description is obtained in the background-enhanced samples.

\begin{table}[t]
    \centering
    \caption{\small Reference branching fractions ($\mathcal{B}$) used for the resonant channels alongside signal yields.}
    \begin{tabular}{llr}
\hline
   Channel               & $\mathcal{B}$ \cite{PDG2018} & Fitted yield \\
\hline
   \DpTopiphiphiTomumu   & $(1.63\pm0.12)\times 10^{-6}$ &             $\num{18100}\pm340$ \\
   \DpTopiphiphiToee     & $(1.67\pm0.07)\times 10^{-6}$ &  \phantom{00}$\num{2160}\pm180$ \\
   \DspTopiphiphiTomumu  & $(1.33\pm0.10)\times 10^{-5}$ &             $\num{42000}\pm400$ \\
   \DspTopiphiphiToee    & $(1.37\pm0.05)\times 10^{-5}$ &  \phantom{00}$\num{5320}\pm180$ \\
\hline
\end{tabular}

    \label{tab:norm-fit-result}
\end{table}

\begin{figure}[htp]
    \centering
    \begin{subfigure}[t]{0.49\textwidth}
        \includegraphics[width=\textwidth]{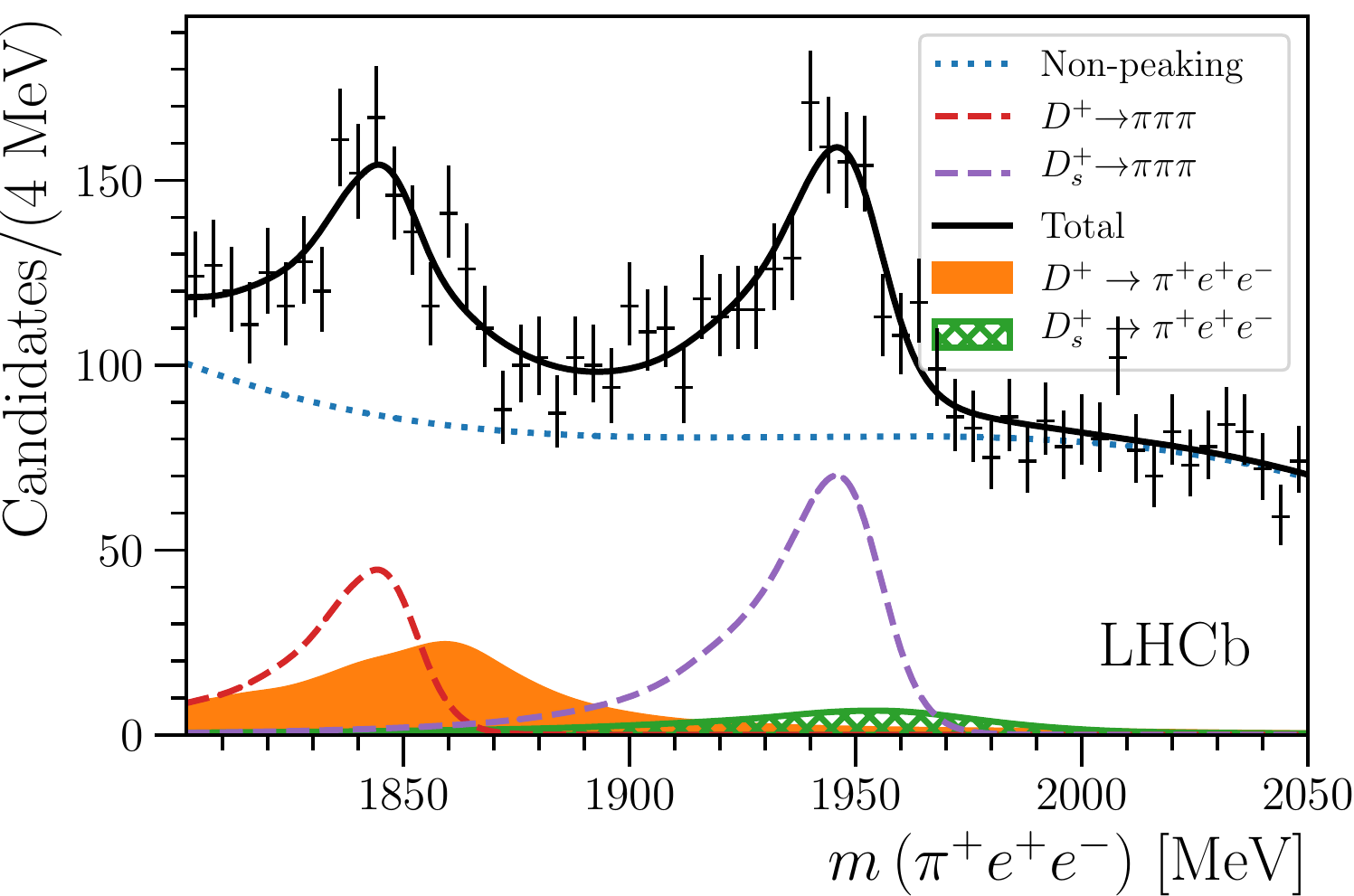}
    \end{subfigure}
    \begin{subfigure}[t]{0.49\textwidth}
        \includegraphics[width=\textwidth]{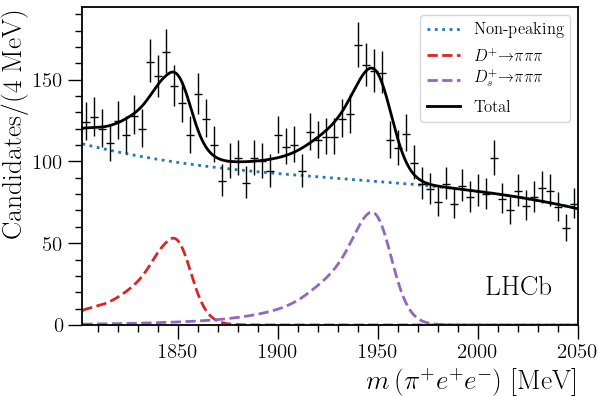}
    \end{subfigure}
    \begin{subfigure}[t]{0.49\textwidth}
        \includegraphics[width=\textwidth]{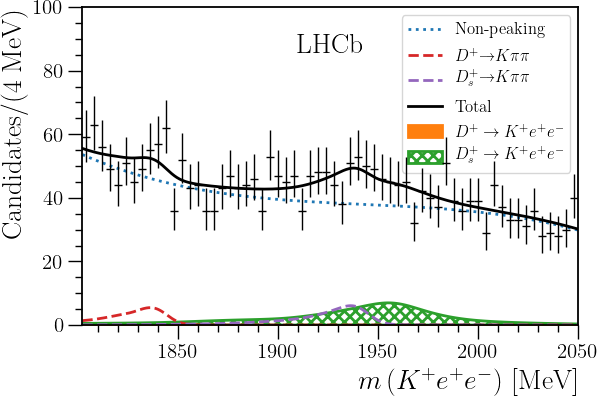}
    \end{subfigure}
    \begin{subfigure}[t]{0.49\textwidth}
        \includegraphics[width=\textwidth]{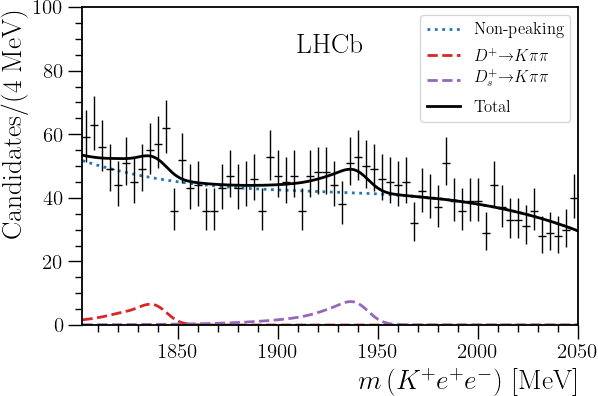}
    \end{subfigure}
    \begin{subfigure}[t]{0.49\textwidth}
        \includegraphics[width=\textwidth]{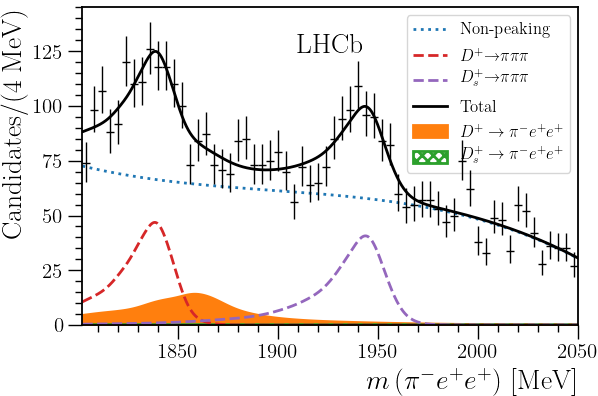}
    \end{subfigure}
    \begin{subfigure}[t]{0.49\textwidth}
        \includegraphics[width=\textwidth]{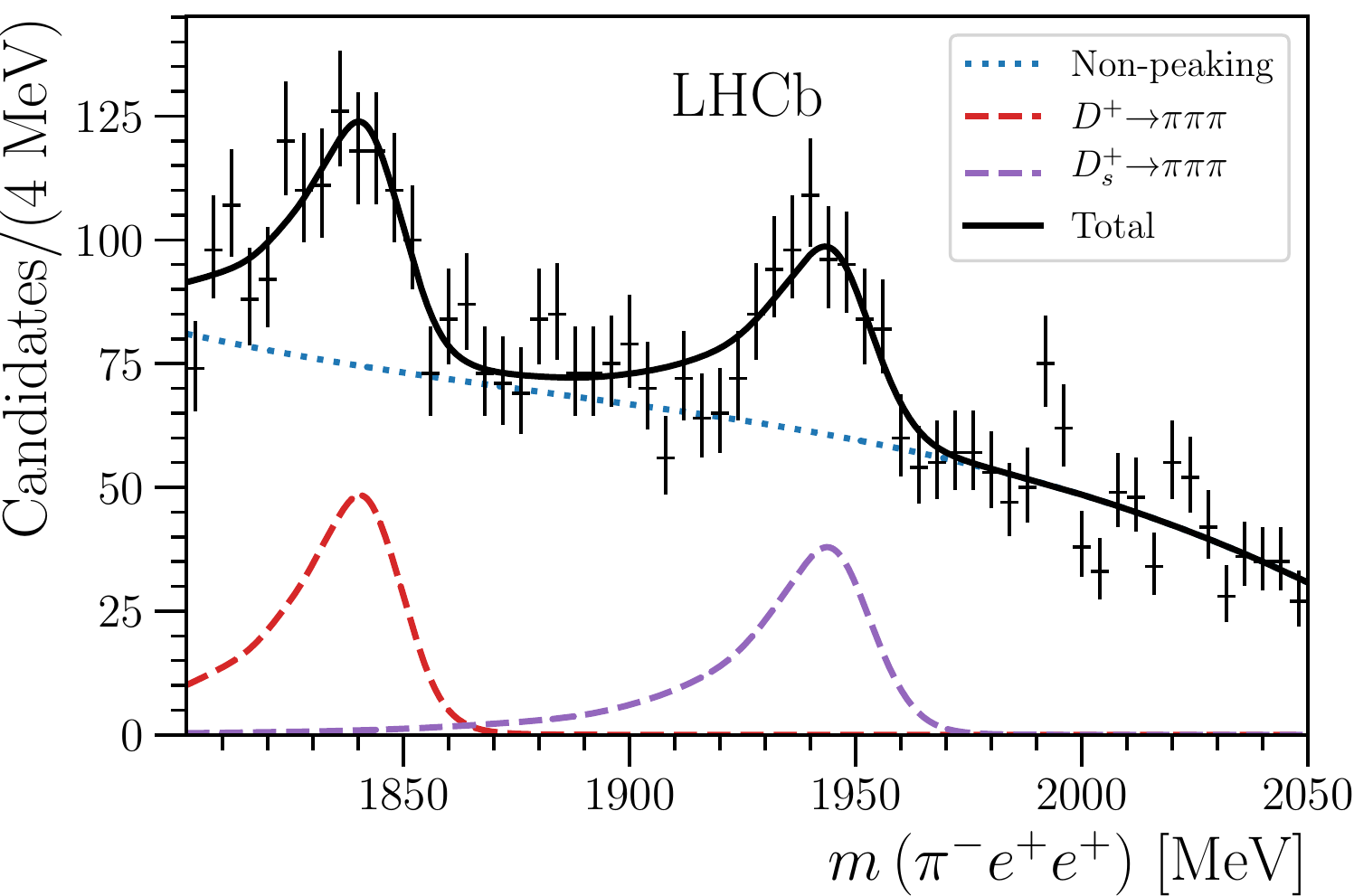}
    \end{subfigure}
    \begin{subfigure}[t]{0.49\textwidth}
        \includegraphics[width=\textwidth]{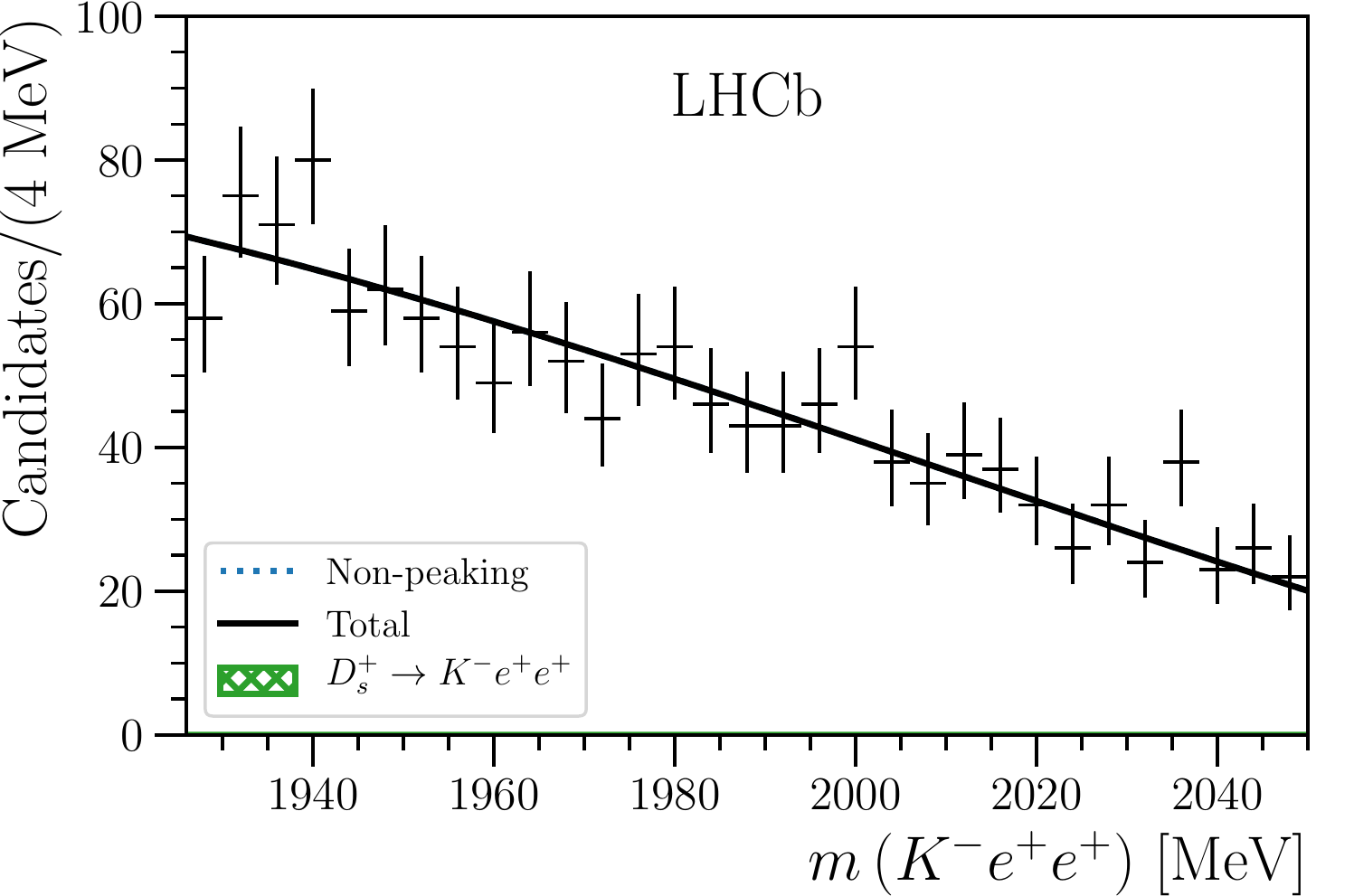}
    \end{subfigure}
    \begin{subfigure}[t]{0.49\textwidth}
        \includegraphics[width=\textwidth]{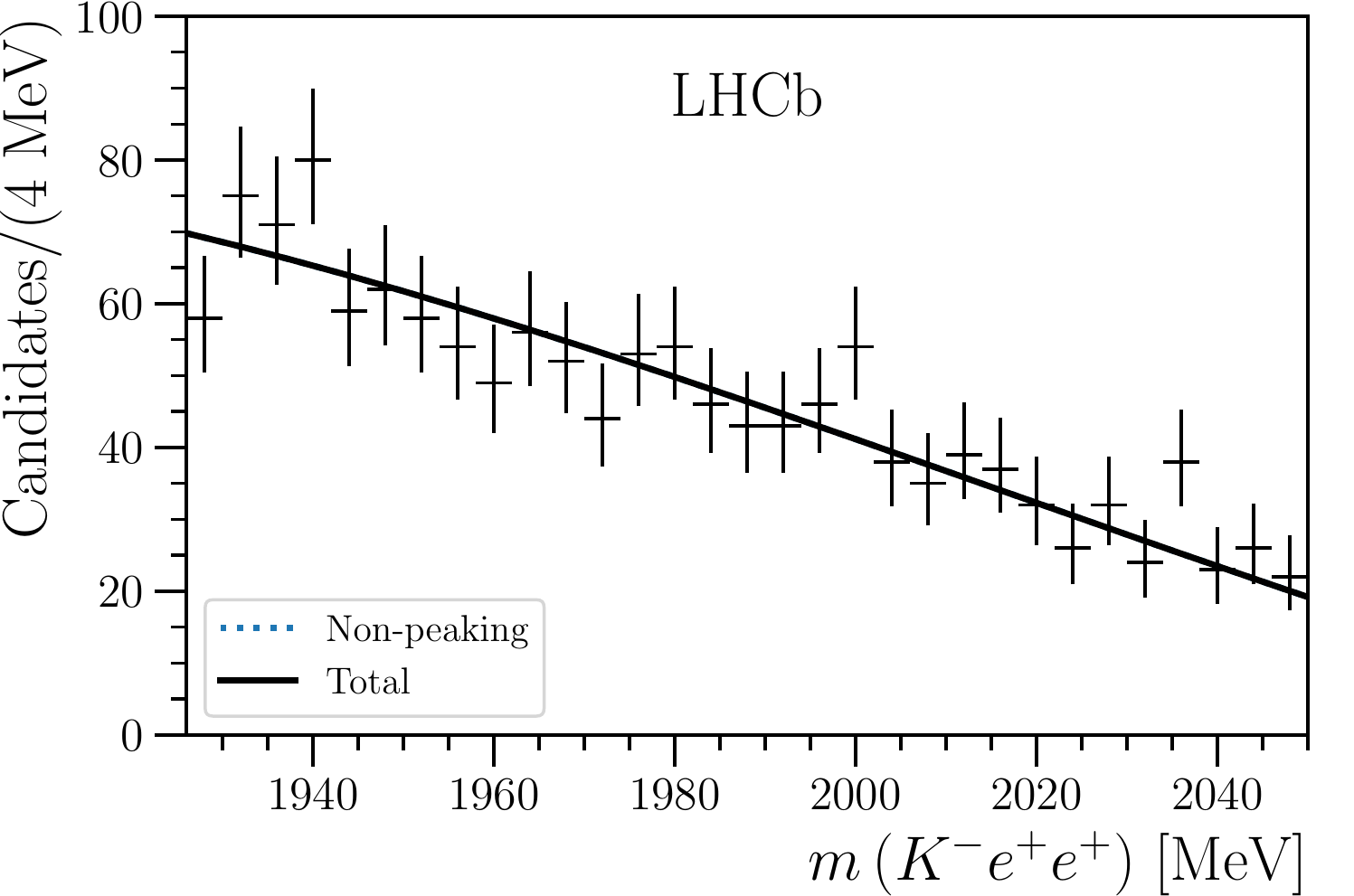}
    \end{subfigure}

    \caption{\small Distributions of the three-body invariant mass in the signal regions for decays with two electrons. The final states are specified in the mass label.
    The left (right) fit with the signal-plus-background (background-only) hypothesis is overlaid with the peaking backgrounds denoted by dashed lines and the non-peaking background is denoted by the blue dotted line.
    }
    \label{fig:mass-fits:3}
\end{figure}

\begin{figure}[htp]
    \centering
    \begin{subfigure}[t]{0.49\textwidth}
        \includegraphics[width=\textwidth]{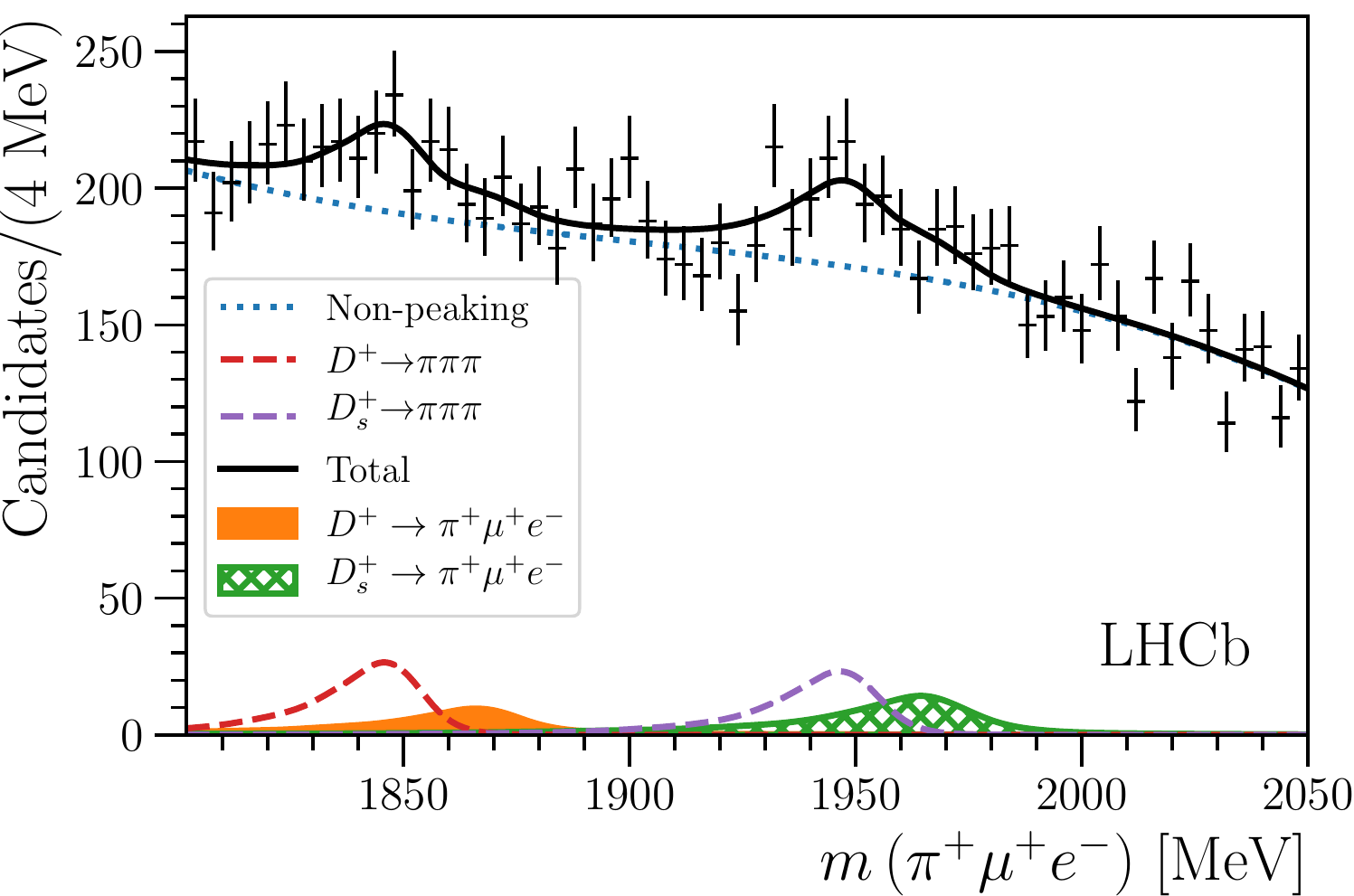}
    \end{subfigure}
    \begin{subfigure}[t]{0.49\textwidth}
        \includegraphics[width=\textwidth]{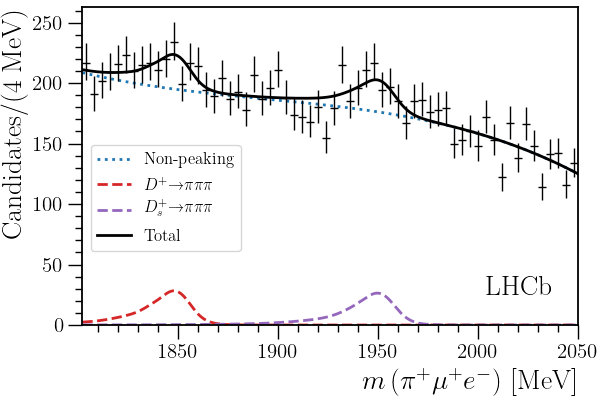}
    \end{subfigure}
    \begin{subfigure}[t]{0.49\textwidth}
        \includegraphics[width=\textwidth]{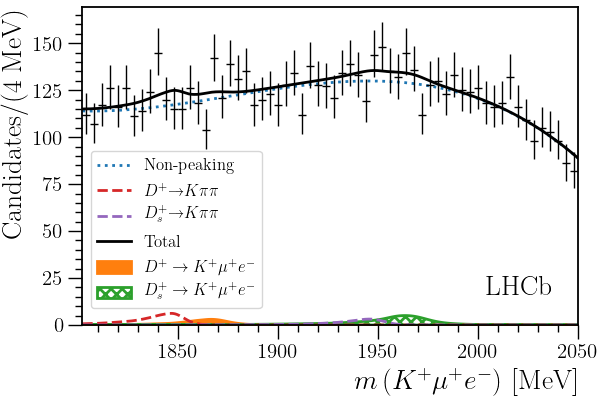}
    \end{subfigure}
    \begin{subfigure}[t]{0.49\textwidth}
        \includegraphics[width=\textwidth]{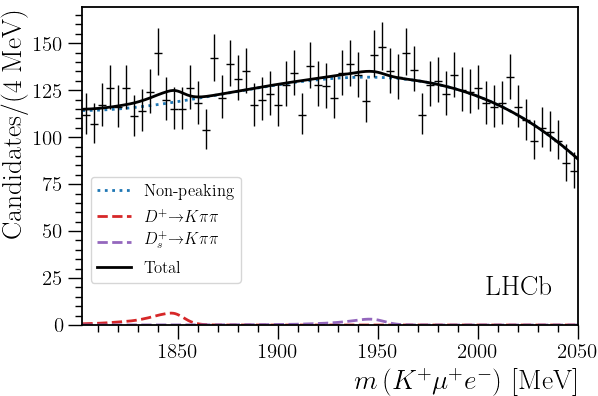}
    \end{subfigure}
    \begin{subfigure}[t]{0.49\textwidth}
        \includegraphics[width=\textwidth]{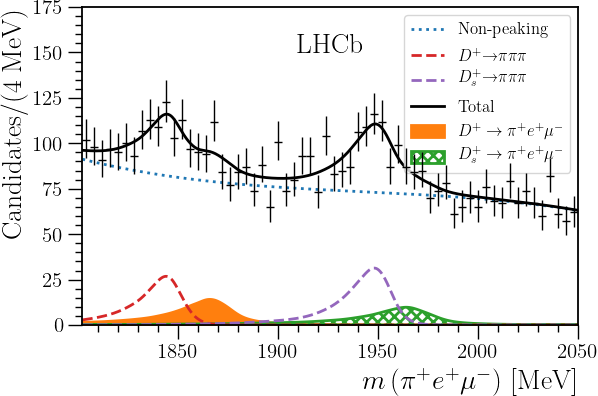}
    \end{subfigure}
    \begin{subfigure}[t]{0.49\textwidth}
        \includegraphics[width=\textwidth]{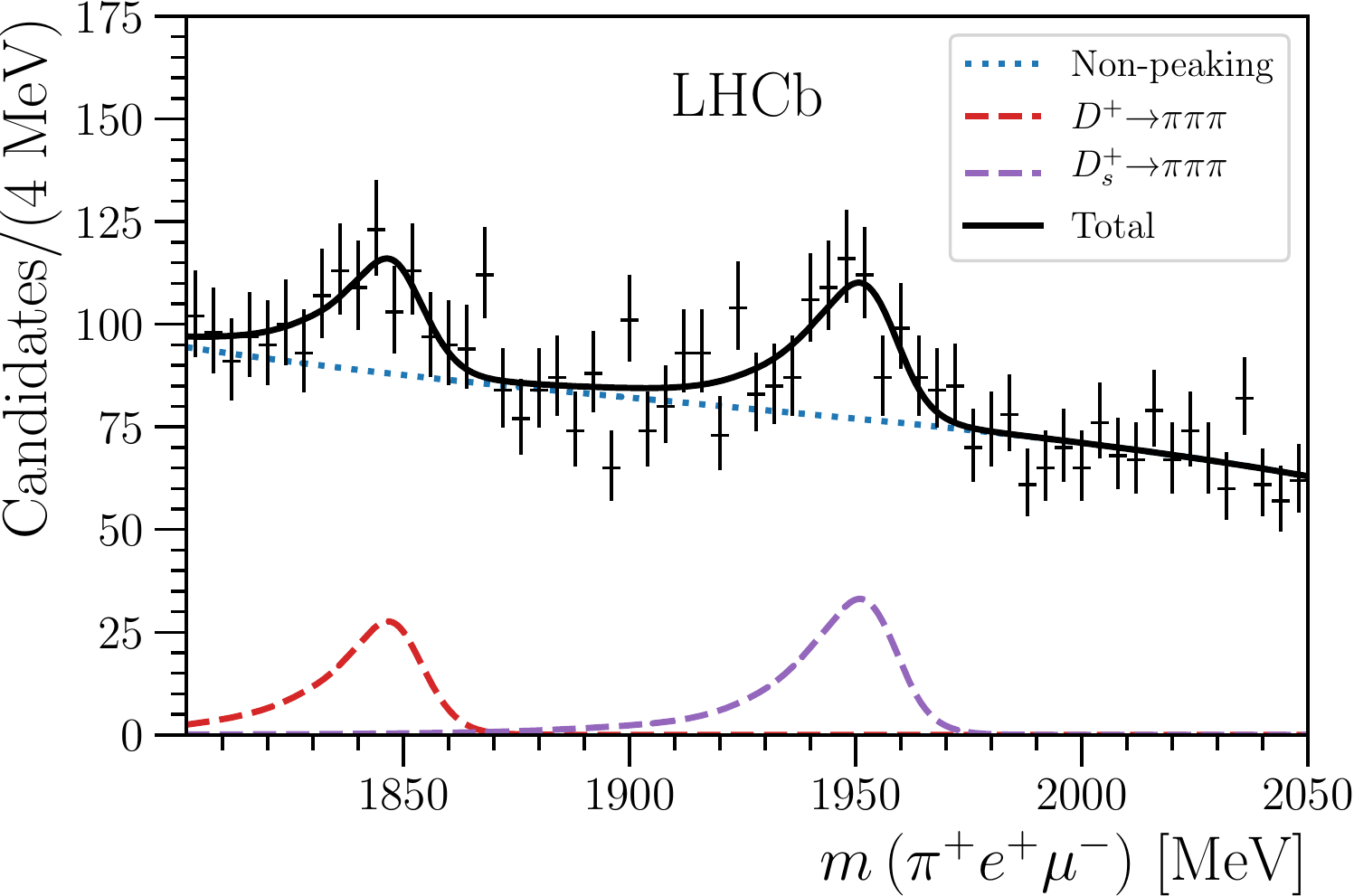}
    \end{subfigure}
    \begin{subfigure}[t]{0.49\textwidth}
        \includegraphics[width=\textwidth]{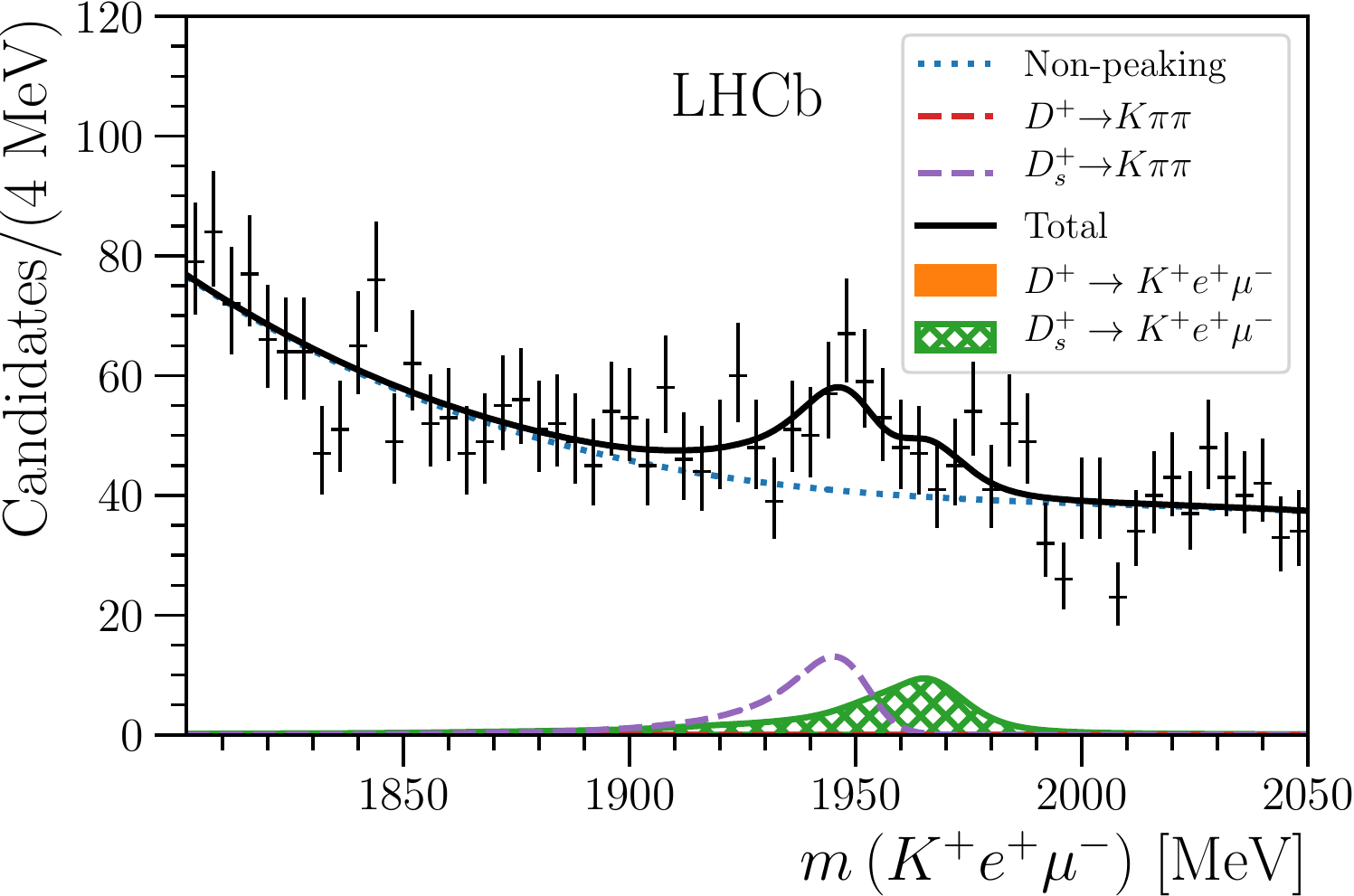}
    \end{subfigure}
    \begin{subfigure}[t]{0.49\textwidth}
        \includegraphics[width=\textwidth]{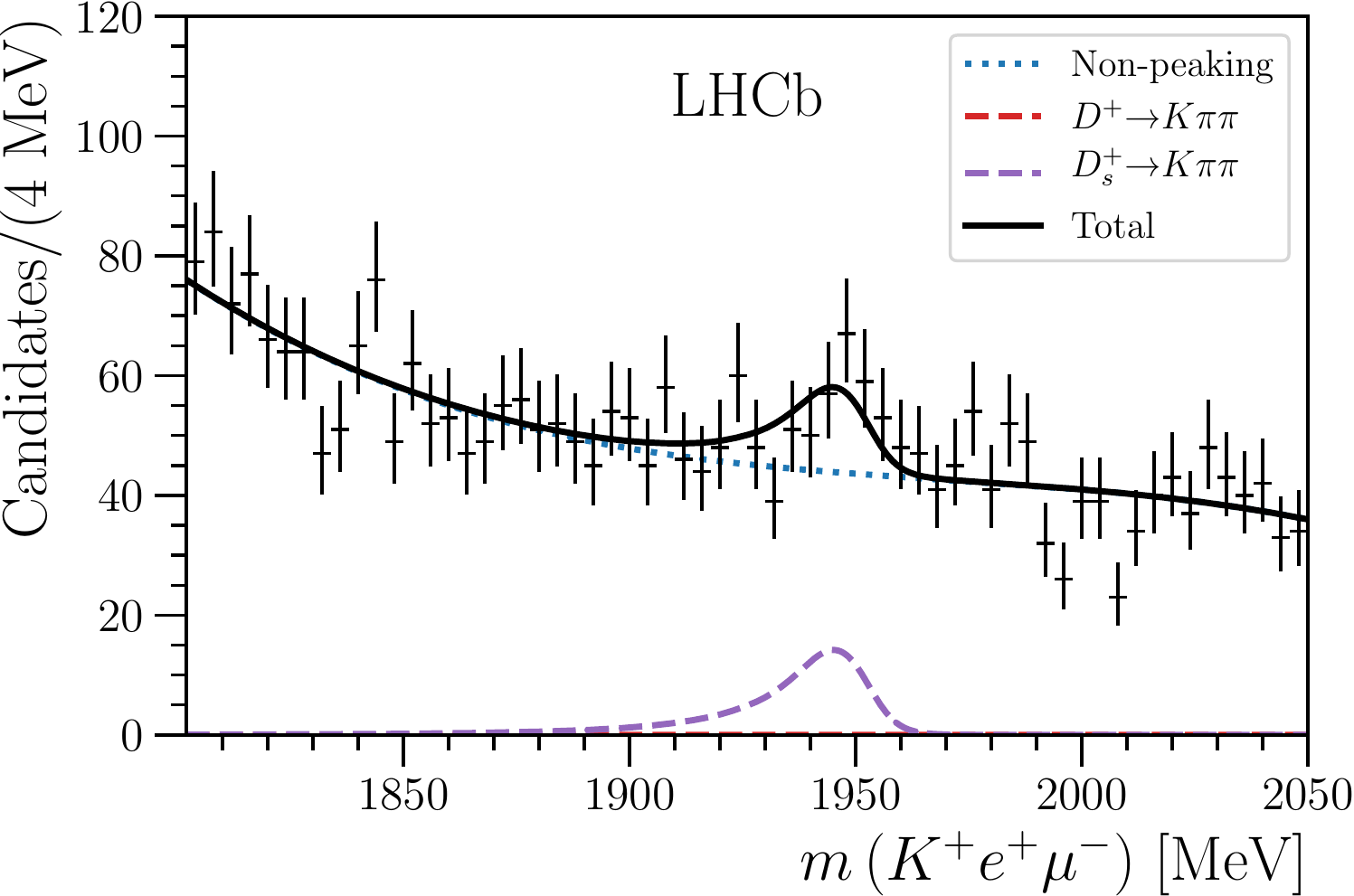}
    \end{subfigure}

    \caption{\small Distributions of the three-body invariant mass in the signal regions for decays with an oppositely charged electron and muon. The final states are specified in the mass label.
    The left (right) fit with the signal-plus-background (background-only) hypothesis is overlaid with the peaking backgrounds denoted by dashed lines and the non-peaking background is denoted by the blue dotted line.
    }
    \label{fig:mass-fits:3}
\end{figure}

\section{Branching fraction determination}
\label{BranchingFraction}

The signal branching fractions are obtained from fits to the three-body invariant-mass distributions, shown in Figs.~\ref{fig:mass-fits:1}-\ref{fig:mass-fits:4}. The branching fractions are defined by
\begin{equation}\label{eqn:br}
\mathcal{B}_{D_{\left(s\right)}^+ \rightarrow h^\pm \ell^+ \ell^{\left(\prime\right) \mp}} =
\frac{N_{ D_{\left(s\right)}^+ \rightarrow h^\pm \ell^+ \ell^{\left(\prime\right) \mp} }}
     {N_{ D_{\left(s\right)}^+ \rightarrow \left(\phi \to \mu^+ \mu^-\right) \pi^+ }}
\cdot
\frac{\epsilon_{ D_{\left(s\right)}^+ \rightarrow \left(\phi \to \mu^+ \mu^-\right) \pi^+ }}
     {\epsilon_{ D_{\left(s\right)}^+ \rightarrow h^\pm \ell^+ \ell^{\left(\prime\right) \mp} }}
\cdot
\mathcal{B}_{ D_{\left(s\right)}^+ \rightarrow \left( \phi \rightarrow \mu^+ \mu^- \right) \pi^+ }
\end{equation}
where $N$ is the fitted yield in the channel indicated, $\mathcal{B}$ is the branching fraction and $\epsilon$ is the efficiency of the selection criteria.
The branching fraction used for \mbox{\DTopiphiphiToll} can be found in Table~\ref{tab:norm-fit-result} and is obtained by combining the measurements of \mbox{\DTopiphiphiToKK}, \mbox{\phiToKK} and \mbox{\phiToll} from Ref.~\cite{PDG2018}.
The systematic uncertainties are described in Section~\ref{SystematicUncertainty} and taken into account with a log-normal distribution in the branching fraction fit.
The potential overlap between the \Dp and \Dsp signal peaks in the same final state is accounted for by floating the yield of the other meson in each fit and treating this as an additional nuisance parameter when computing the significance of the signal peak.

The effect of discrepancies between data and simulation are reduced by using a reweighting technique that utilises a multivariate classifier. This classifier is used to generate per-event weights using the method outlined in Ref.~\cite{rogozhnikov:2016bdp}.
The procedure is applied separately for each of the calibration channels and each classifier is trained to distinguish between real and simulated calibration-channel events.
Background subtraction of the data is applied using the \sPlot technique~\cite{Pivk:2004ty} with a fit to the invariant mass of the calibration channel.
The weights obtained from these classifiers are applied to the simulation and then used throughout this analysis.
Decay channels with two electrons in the final state use weights from the \mbox{\DTopiphiphiToee} classifiers and all other decays use the \mbox{\DTopiphiphiTomumu} classifiers.
In all cases the same parent particle (\DorDsp) decay as the signal is used.

Equation~(\ref{eqn:br}) includes the ratio of efficiencies between the normalisation channel and the signal channel so that the effects of several systematic uncertainties cancel.
The reconstruction and selection efficiencies are obtained from simulation with the aforementioned corrections applied.
As the resonant structure in the signal channels is inherently unknown, the efficiency of each signal decay is obtained under the assumption that the decay particles are uniformly distributed across the phase space with the signal contribution extrapolated into any vetoed kinematic regions.

The efficiency with which events pass the PID requirements is obtained using a set of calibration channels. The PID response is sampled~\cite{meerkat} as a function of the particle kinematics and event multiplicity in simulation, as described in Ref.~\cite{LHCb-DP-2018-001}.
A small correction is made for differences in the track reconstruction efficiency between data and simulation~\cite{LHCb-DP-2013-002}.

For channels containing electrons, an additional correction is applied to the ratio of electron to muon reconstruction efficiencies.
This is computed under the assumption that the correction for each electron is independent of the other, and thus the square root of the efficiency-corrected signal-yield ratio between the calibration channels \mbox{\DTopiphiphiToee} and \mbox{\DTopiphiphiTomumu} in data is taken.

The single event sensitivities, i.e. the branching fractions corresponding to a single observed signal event, vary from \num{3e-10} to \num{1e-8} depending on the channel.
To ensure the efficiencies are sufficiently well understood, cross-check measurements are made of the branching fractions of \mbox{\DTopiphiphiTomumu} and \mbox{\DTopiphiphiToee} decays.
The results obtained before and after applying the offline selection criteria are compared and found to be in agreement.

\section{Systematic uncertainties}
\label{SystematicUncertainty}

The efficiency ratio in Eq.~\ref{eqn:br} is affected by several sources of systematic uncertainty.
The finite size of the simulated signal event samples introduces a systematic uncertainty on the branching fractions varying from \SIrange{1.4}{6.4}{\percent}.
A summary of the systematic uncertainties assigned for each effect is given in Table~\ref{tab:systematics}.

The track-reconstruction efficiency from simulation is corrected using a tag-and-probe technique in data where \JPsiTomumu decays are selected by making requirements on only one muon.
This is found to have negligible effect on the efficiency ratio.
A \SI{0.8}{\percent} systematic uncertainty is assigned for each particle species in the signal decay that differs from that of the \mbox{\DTopiphiphiTomumu} reference.
An additional \SI{1.5}{\percent} systematic uncertainty is assigned to channels containing a kaon to account for possible hadronic interaction effects with the detector material~\cite{LHCb-DP-2013-002}.
Alternative parametrisations~\cite{LHCb-DP-2018-001} are considered for the sampling of the PID response, but their associated systematic effects have negligible impact compared to the statistical uncertainties of the analysis and other sources of systematic uncertainty.
For final states with both a muon and an electron, a further systematic uncertainty of \SI{7.6}{\percent} is assigned for the choice of reweighting classifier.
This is obtained from the RMS of the change in efficiency of all mixed-lepton final states, using the alternative reweighting classifier.

The systematic uncertainties affecting the signal yield are estimated using an alternative fit model for the normalisation channels.
For signal states with two muons, a KDE parametrisation obtained from simulation is used as an alternative for the nominal model describing the \DTopiphiphiTomumu signal.
In all other cases, the differences are most likely to arise from the treatment of bremsstrahlung radiation in simulation.
Therefore, an alternative model is generated by fixing the relative yield between the three bremsstrahlung categories of \DTopiphiphiToee from simulation.
The signal yield from the alternative models is compared with the nominal model and the difference is assigned as a systematic uncertainty.

The uncertainty on the yield of \DTopiphiphiTomumu is dominated by the modelling of the \DTopipipi backgrounds in the lower tails of the larger signal components.
To account for this, the data set is refitted neglecting these background components,
and the change in signal yield is assigned as a systematic uncertainty.
The same procedure is applied for \DTopiphiphiToee with the change in signal yield used as the uncertainty on the electron efficiency correction factor.

\begin{sidewaystable}[p!]
    \centering
    \small
    \caption{Summary of systematic uncertainties for each signal decay. The ``\phiTomumu yield'' column represents the combination of the statistical uncertainty from the \DTopiphiphiTomumu dataset with the systematic uncertainty from the modelling of the \DTopipipi backgrounds. A further \SI{7.6}{\percent} uncertainty from the branching fraction of \DTopiphiphiTomumu and a \SI{1.0}{\percent} uncertainty from the finite size of the simulated \DTopiphiphiTomumu sample apply to all channels. The electron efficiency correction, described at the end of Section~\ref{BranchingFraction}, is denoted by $\epsilon_{\text{electron}}$. All values are given in percent as a fractional uncertainty on the signal yield.}
    \label{tab:systematics}
    \begin{tabular}{lcccccc}
\toprule
 \multirow{2}{*}{Channel}              & Simulated signal  & Track-reconstruction &\multirow{2}{*}{ Reweighting $\left[\%\right]$}   & \multirow{2}{*}{Signal $\left[\%\right]$} & \multirow{2}{*}{\phiTomumu yield $\left[\%\right]$} & \multirow{2}{*}{\phiToee $\epsilon_\text{electron}$ $\left[\%\right]$} \\
 & sample $\left[\%\right]$ & efficiency $\left[\%\right]$ &                &     &                                   &                                                      \\
\midrule
 $D^+ \rightarrow K^+ e^+ e^-$         & \num{6.4}         & \num{3.2}            & --                       & \num{7.0}          & \num{2.0}                         & \num{9.1}                  \\
 $D^+ \rightarrow K^+ e^+ \mu^-$       & \num{2.1}         & \num{2.5}            & \num{7.6}                & \num{7.0}          & \num{2.0}                         & \num{4.6}                  \\
 $D^+ \rightarrow K^+ \mu^+ e^-$       & \num{1.8}         & \num{2.5}            & \num{7.6}                & \num{7.0}          & \num{2.0}                         & \num{4.6}                  \\
 $D^+ \rightarrow K^+ \mu^+ \mu^-$     & \num{3.0}         & \num{2.3}            & --                       & \num{0.8}          & \num{2.0}                         & --                         \\
 $D^+ \rightarrow \pi^+ e^+ e^-$       & \num{3.9}         & \num{2.3}            & --                       & \num{7.0}          & \num{2.0}                         & \num{9.1}                  \\
 $D^+ \rightarrow \pi^- e^+ e^+$       & \num{2.6}         & \num{2.3}            & --                       & \num{7.0}          & \num{2.0}                         & \num{9.1}                  \\
 $D^+ \rightarrow \pi^+ e^+ \mu^-$     & \num{1.9}         & \num{1.1}            & \num{7.6}                & \num{7.0}          & \num{2.0}                         & \num{4.6}                  \\
 $D^+ \rightarrow \pi^+ \mu^+ e^-$     & \num{1.5}         & \num{1.1}            & \num{7.6}                & \num{7.0}          & \num{2.0}                         & \num{4.6}                  \\
 $D^+ \rightarrow \pi^- \mu^+ e^+$     & \num{1.6}         & \num{1.1}            & \num{7.6}                & \num{7.0}          & \num{2.0}                         & \num{4.6}                  \\
 $D^+ \rightarrow \pi^+ \mu^+ \mu^-$   & \num{1.9}         & \num{0.1}            & --                       & \num{0.8}          & \num{2.0}                         & --                         \\
 $D^+ \rightarrow \pi^- \mu^+ \mu^+$   & \num{1.5}         & \num{0.0}            & --                       & \num{0.8}          & \num{2.0}                         & --                         \\
\midrule
 $D_s^+ \rightarrow K^+ e^+ e^-$       & \num{6.0}         & \num{3.2}            & --                       & \num{7.0}          & \num{3.0}                         & \num{7.5}                  \\
 $D_s^+ \rightarrow K^- e^+ e^+$       & \num{3.2}         & \num{3.2}            & --                       & \num{7.0}          & \num{3.0}                         & \num{7.5}                  \\
 $D_s^+ \rightarrow K^+ e^+ \mu^-$     & \num{3.0}         & \num{2.5}            & \num{7.6}                & \num{7.0}          & \num{3.0}                         & \num{3.8}                  \\
 $D_s^+ \rightarrow K^+ \mu^+ e^-$     & \num{2.5}         & \num{2.5}            & \num{7.6}                & \num{7.0}          & \num{3.0}                         & \num{3.8}                  \\
 $D_s^+ \rightarrow K^- \mu^+ e^+$     & \num{2.4}         & \num{2.5}            & \num{7.6}                & \num{7.0}          & \num{3.0}                         & \num{3.8}                  \\
 $D_s^+ \rightarrow K^+ \mu^+ \mu^-$   & \num{3.3}         & \num{2.3}            & --                       & \num{0.8}          & \num{3.0}                         & --                         \\
 $D_s^+ \rightarrow K^- \mu^+ \mu^+$   & \num{1.8}         & \num{2.3}            & --                       & \num{0.8}          & \num{3.0}                         & --                         \\
 $D_s^+ \rightarrow \pi^+ e^+ e^-$     & \num{4.3}         & \num{2.3}            & --                       & \num{7.0}          & \num{3.0}                         & \num{7.5}                  \\
 $D_s^+ \rightarrow \pi^- e^+ e^+$     & \num{3.1}         & \num{2.3}            & --                       & \num{7.0}          & \num{3.0}                         & \num{7.5}                  \\
 $D_s^+ \rightarrow \pi^+ e^+ \mu^-$   & \num{2.7}         & \num{1.2}            & \num{7.6}                & \num{7.0}          & \num{3.0}                         & \num{3.8}                  \\
 $D_s^+ \rightarrow \pi^+ \mu^+ e^-$   & \num{2.3}         & \num{1.1}            & \num{7.6}                & \num{7.0}          & \num{3.0}                         & \num{3.8}                  \\
 $D_s^+ \rightarrow \pi^- \mu^+ e^+$   & \num{2.2}         & \num{1.2}            & \num{7.6}                & \num{7.0}          & \num{3.0}                         & \num{3.8}                  \\
 $D_s^+ \rightarrow \pi^+ \mu^+ \mu^-$ & \num{2.5}         & \num{0.1}            & --                       & \num{0.8}          & \num{3.0}                         & --                         \\
 $D_s^+ \rightarrow \pi^- \mu^+ \mu^+$ & \num{2.2}         & \num{0.0}            & --                       & \num{0.8}          & \num{3.0}                         & --                         \\
\bottomrule
\end{tabular}

\end{sidewaystable}

\section{Results and conclusions}
\label{results}

\begin{figure}[p]
    \centering
    \begin{subfigure}[t]{0.8\textwidth}
        \centering
        \includegraphics[width=\textwidth]{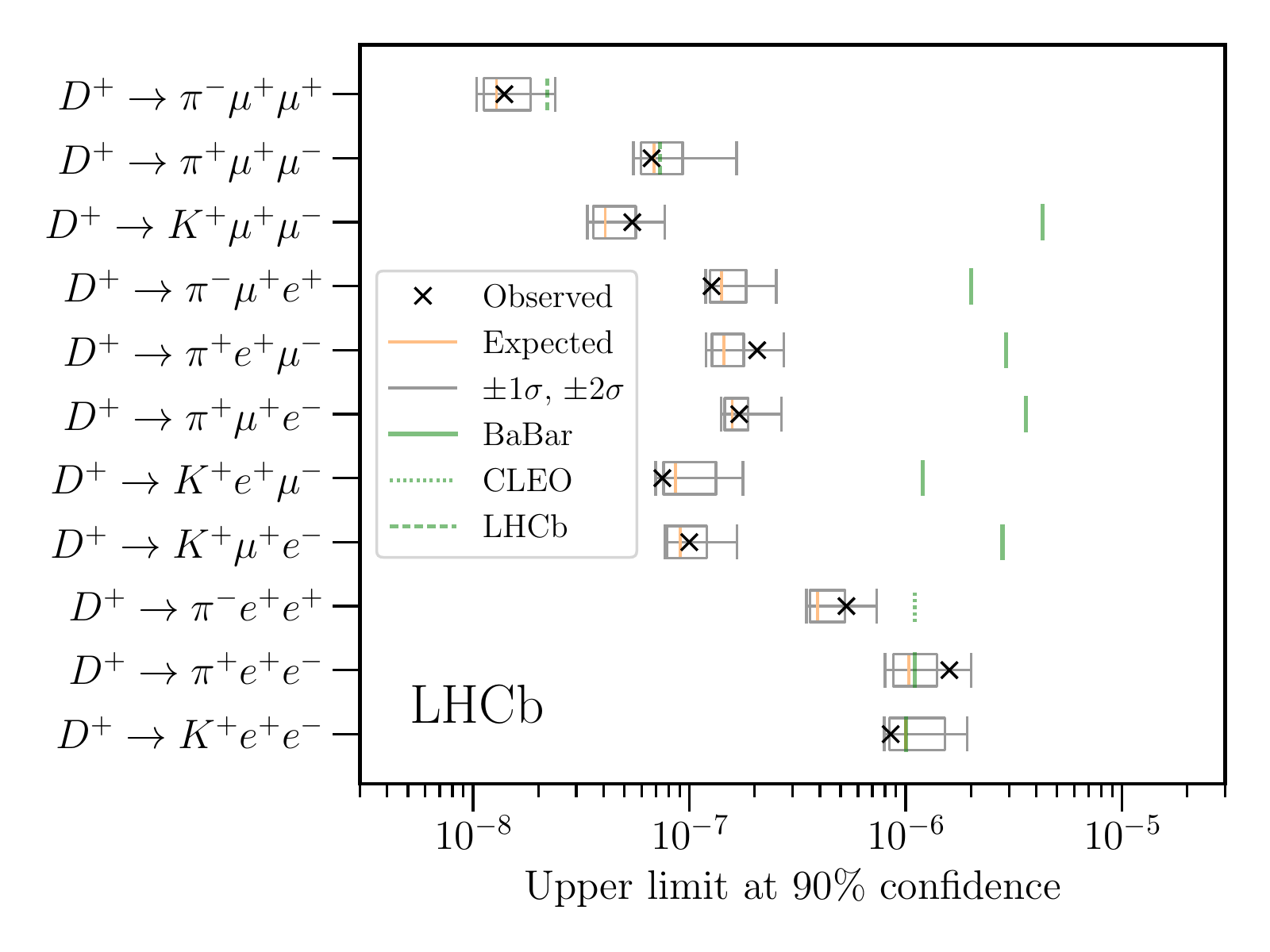}
    \end{subfigure}%

    \begin{subfigure}[t]{0.8\textwidth}
        \centering
        \includegraphics[width=\textwidth]{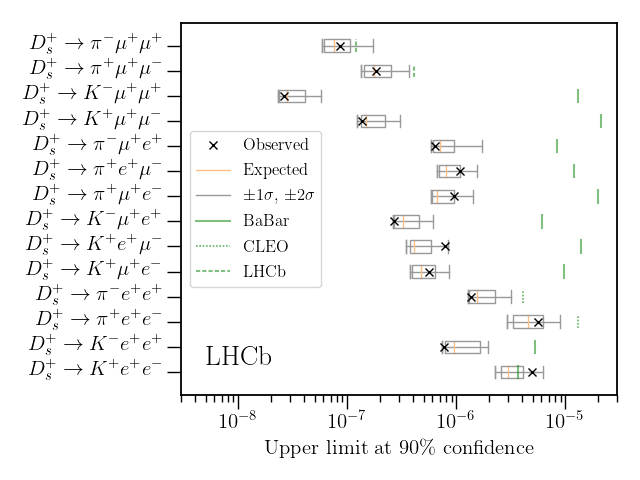}
    \end{subfigure}%

    \caption{\small Upper limits at \SI{90}{\percent} confidence level on the \DorDsp signal channels. The median (orange), $\pm1\sigma$ and $\pm2\sigma$ expected limits are shown as box plots and the observed limit is given by a black cross. The green line shows the previous world's best limit for each channel where the solid, dotted and dashed lines correspond to BaBar, CLEO and LHCb~\cite{Lees:2011hb,Rubin:2010cq,LHCB-PAPER-2012-051}.}
    \label{fig:2016Results}
\end{figure}

No significant deviation from the background only hypothesis is found for any of the channels and all limits are within $\pm2\sigma$ of the expected limit.
The compatibility of the observed mass-distributions with a signal-plus-background or a background-only hypothesis is evaluated using the CL$_s$ method
with systematic uncertainties included as described in Ref.~\cite{CLs,Junk:1999kv}.
Upper limits on the branching fractions are determined using the observed distribution of CL$_s$ as a function of the assumed branching fraction.
The upper limits at \SI{90}{\percent} and \SI{95}{\percent} confidence level (CL) are given in Table~\ref{tab:2016Results} and are shown in Fig.~\ref{fig:2016Results}.

\begin{table}[t]
    \centering
    \caption{The single event sensitivities (SES), and upper limits on the branching fractions obtained using the CL$_s$ method, for each signal decay channel.}

    \begin{tabular}{@{}l|rrr|rrr@{}}
        \multirow{3}{*}{Decay} & \multicolumn{4}{c}{Branching fraction upper limit [$10^{-9}$]} \\
         & \multicolumn{3}{c|}{\Dp} & \multicolumn{3}{c}{\Dsp} \\
         & SES & \SI{90}{\percent} CL & \SI{95}{\percent} CL
         & SES & \SI{90}{\percent} CL & \SI{95}{\percent} CL \\
        \midrule
        \DTopimumuOS & 0.6 & 67 & 74 & 2.4 & 180 & 210 \\
        \DTopimumuSS & 0.3 & 14 & 16 & 1.8 & 86 & 96 \\
        \DToKmumuOS & 1.2 & 54 & 61 & 3.8 & 140 & 160 \\
        \DToKmumuSS & - & - & - & 1.2 & 26 & 30 \\
        \DTopiemuOS & 0.6 & 210 & 230 & 3.1 & 1100 & 1200 \\
        \DTopimueOS & 0.4 & 220 & 220 & 2.2 & 940 & 1100 \\
        \DTopimueSS & 0.4 & 130 & 150 & 2.0 & 630 & 710 \\
        \DToKemuOS & 0.7 & 75 & 83 & 3.7 & 790 & 880 \\
        \DToKmueOS & 0.5 & 100 & 110 & 2.5 & 560 & 640 \\
        \DToKmueSS & - & - & - & 2.4 & 260 & 320 \\
        \DTopieeOS & 1.9 & 1600 & 1800 & 8.1 & 5500 & 6400 \\
        \DTopieeSS & 0.9 & 530 & 600 & 4.1 & 1400 & 1600 \\
        \DToKeeOS & 4.4 & 850 & 1000 & 14.8 & 4900 & 5500 \\
        \DToKeeSS & - & - & - & 4.1 & 770 & 840 \\
    \end{tabular}

    \label{tab:2016Results}
\end{table}

In conclusion, searches have been made for 25 previously unobserved three-body decays of \Dp and \Dsp mesons using \SI{1.6}{\per\femto\barn} of \proton\proton collision data collected by the LHCb experiment during 2016.
The decays are of the form \DTohll where $\Ph$ is a kaon or pion and $\ell^{(\prime)}$ is an electron or muon.
No significant deviations from the background-only hypotheses are seen.
The \SI{90}{\percent} CL limits on the branching fractions vary from \num{1.4e-8} to \num{6.4e-6}
and represent the world's best limits for 23 of these decays. In the majority of the channels the improvement on previous limits is more than an order of magnitude, with the largest improvement being a factor of five hundred in \DspToKmumuSS decays.

\section*{Acknowledgements}
\noindent We express our gratitude to our colleagues in the CERN
accelerator departments for the excellent performance of the LHC. We
thank the technical and administrative staff at the LHCb
institutes.
We acknowledge support from CERN and from the national agencies:
CAPES, CNPq, FAPERJ and FINEP (Brazil); 
MOST and NSFC (China); 
CNRS/IN2P3 (France); 
BMBF, DFG and MPG (Germany); 
INFN (Italy); 
NWO (Netherlands); 
MNiSW and NCN (Poland); 
MEN/IFA (Romania); 
MSHE (Russia); 
MICINN (Spain); 
SNSF and SER (Switzerland); 
NASU (Ukraine); 
STFC (United Kingdom); 
DOE NP and NSF (USA).
We acknowledge the computing resources that are provided by CERN, IN2P3
(France), KIT and DESY (Germany), INFN (Italy), SURF (Netherlands),
PIC (Spain), GridPP (United Kingdom), RRCKI and Yandex
LLC (Russia), CSCS (Switzerland), IFIN-HH (Romania), CBPF (Brazil),
PL-GRID (Poland) and OSC (USA).
We are indebted to the communities behind the multiple open-source
software packages on which we depend.
Individual groups or members have received support from
AvH Foundation (Germany);
EPLANET, Marie Sk\l{}odowska-Curie Actions and ERC (European Union);
A*MIDEX, ANR, Labex P2IO and OCEVU, and R\'{e}gion Auvergne-Rh\^{o}ne-Alpes (France);
Key Research Program of Frontier Sciences of CAS, CAS PIFI,
Thousand Talents Program, and Sci. \& Tech. Program of Guangzhou (China);
RFBR, RSF and Yandex LLC (Russia);
GVA, XuntaGal and GENCAT (Spain);
the Royal Society
and the Leverhulme Trust (United Kingdom).

\addcontentsline{toc}{section}{References}
\bibliographystyle{template/LHCb}
\bibliography{main,template/bib/standard,template/bib/LHCb-PAPER,template/bib/LHCb-CONF,template/bib/LHCb-DP,template/bib/LHCb-TDR}

\ifx\mcitethebibliography\mciteundefinedmacro
\PackageError{LHCb.bst}{mciteplus.sty has not been loaded}
{This bibstyle requires the use of the mciteplus package.}\fi
\providecommand{\href}[2]{#2}
\begin{mcitethebibliography}{10}
\mciteSetBstSublistMode{n}
\mciteSetBstMaxWidthForm{subitem}{\alph{mcitesubitemcount})}
\mciteSetBstSublistLabelBeginEnd{\mcitemaxwidthsubitemform\space}
{\relax}{\relax}

\bibitem{Glashow:1970gm}
S.~L. Glashow, J.~Iliopoulos, and L.~Maiani,
  \ifthenelse{\boolean{articletitles}}{\emph{{Weak interactions with
  lepton-hadron symmetry}},
  }{}\href{https://doi.org/10.1103/PhysRevD.2.1285}{Phys.\ Rev.\  \textbf{D2}
  (1970) 1285}\relax
\mciteBstWouldAddEndPuncttrue
\mciteSetBstMidEndSepPunct{\mcitedefaultmidpunct}
{\mcitedefaultendpunct}{\mcitedefaultseppunct}\relax
\EndOfBibitem
\bibitem{PDG2018}
Particle Data Group, M.~Tanabashi {\em et~al.},
  \ifthenelse{\boolean{articletitles}}{\emph{{\href{http://pdg.lbl.gov/}{Review
  of particle physics}}},
  }{}\href{https://doi.org/10.1103/PhysRevD.98.030001}{Phys.\ Rev.\
  \textbf{D98} (2018) 030001}\relax
\mciteBstWouldAddEndPuncttrue
\mciteSetBstMidEndSepPunct{\mcitedefaultmidpunct}
{\mcitedefaultendpunct}{\mcitedefaultseppunct}\relax
\EndOfBibitem
\bibitem{LHCb-PAPER-2014-049}
CMS and LHCb collaborations, V.~Khachatryan {\em et~al.},
  \ifthenelse{\boolean{articletitles}}{\emph{{Observation of the rare
  \mbox{\decay{\Bs}{\mumu}} decay from the combined analysis of CMS and LHCb
  data}}, }{}\href{https://doi.org/10.1038/nature14474}{Nature \textbf{522}
  (2015) 68}, \href{http://arxiv.org/abs/1411.4413}{{\normalfont\ttfamily
  arXiv:1411.4413}}\relax
\mciteBstWouldAddEndPuncttrue
\mciteSetBstMidEndSepPunct{\mcitedefaultmidpunct}
{\mcitedefaultendpunct}{\mcitedefaultseppunct}\relax
\EndOfBibitem
\bibitem{LHCb-PAPER-2017-001}
LHCb collaboration, R.~Aaij {\em et~al.},
  \ifthenelse{\boolean{articletitles}}{\emph{{Measurement of the
  \mbox{\decay{\Bs}{\mumu}} branching fraction and effective lifetime and
  search for \mbox{\decay{\Bz}{\mumu}} decays}},
  }{}\href{https://doi.org/10.1103/PhysRevLett.118.191801}{Phys.\ Rev.\ Lett.\
  \textbf{118} (2017) 191801},
  \href{http://arxiv.org/abs/1703.05747}{{\normalfont\ttfamily
  arXiv:1703.05747}}\relax
\mciteBstWouldAddEndPuncttrue
\mciteSetBstMidEndSepPunct{\mcitedefaultmidpunct}
{\mcitedefaultendpunct}{\mcitedefaultseppunct}\relax
\EndOfBibitem
\bibitem{LHCb-PAPER-2019-009}
LHCb collaboration, R.~Aaij {\em et~al.},
  \ifthenelse{\boolean{articletitles}}{\emph{{Search for lepton-universality
  violation in \mbox{\decay{\Bp}{\Kp\ellell}} decays}},
  }{}\href{https://doi.org/10.1103/PhysRevLett.122.191801}{Phys.\ Rev.\ Lett.\
  \textbf{122} (2019) 191801},
  \href{http://arxiv.org/abs/1903.09252}{{\normalfont\ttfamily
  arXiv:1903.09252}}\relax
\mciteBstWouldAddEndPuncttrue
\mciteSetBstMidEndSepPunct{\mcitedefaultmidpunct}
{\mcitedefaultendpunct}{\mcitedefaultseppunct}\relax
\EndOfBibitem
\bibitem{deBoer:2015boa}
S.~de~Boer and G.~Hiller, \ifthenelse{\boolean{articletitles}}{\emph{{Flavor
  and new physics opportunities with rare charm decays into leptons}},
  }{}\href{https://doi.org/10.1103/PhysRevD.93.074001}{Phys.\ Rev.\
  \textbf{D93} (2016) 074001},
  \href{http://arxiv.org/abs/1510.00311}{{\normalfont\ttfamily
  arXiv:1510.00311}}\relax
\mciteBstWouldAddEndPuncttrue
\mciteSetBstMidEndSepPunct{\mcitedefaultmidpunct}
{\mcitedefaultendpunct}{\mcitedefaultseppunct}\relax
\EndOfBibitem
\bibitem{Fajfer:2015mia}
S.~Fajfer and N.~Košnik, \ifthenelse{\boolean{articletitles}}{\emph{{Prospects
  of discovering new physics in rare charm decays}},
  }{}\href{https://doi.org/10.1140/epjc/s10052-015-3801-2}{Eur.\ Phys.\ J.\
  \textbf{C75} (2015) 567},
  \href{http://arxiv.org/abs/1510.00965}{{\normalfont\ttfamily
  arXiv:1510.00965}}\relax
\mciteBstWouldAddEndPuncttrue
\mciteSetBstMidEndSepPunct{\mcitedefaultmidpunct}
{\mcitedefaultendpunct}{\mcitedefaultseppunct}\relax
\EndOfBibitem
\bibitem{Bause:2019vpr}
R.~Bause, M.~Golz, G.~Hiller, and A.~Tayduganov,
  \ifthenelse{\boolean{articletitles}}{\emph{{The new physics reach of null
  tests with $D \rightarrow \pi \ell \ell $ and $D_s \rightarrow K \ell \ell $
  decays}}, }{}\href{https://doi.org/10.1140/epjc/s10052-020-7621-7}{Eur.\
  Phys.\ J.\ C \textbf{80} (2020) 65},
  \href{http://arxiv.org/abs/1909.11108}{{\normalfont\ttfamily
  arXiv:1909.11108}}\relax
\mciteBstWouldAddEndPuncttrue
\mciteSetBstMidEndSepPunct{\mcitedefaultmidpunct}
{\mcitedefaultendpunct}{\mcitedefaultseppunct}\relax
\EndOfBibitem
\bibitem{Bause:2020obd}
R.~Bause, H.~Gisbert, M.~Golz, and G.~Hiller,
  \ifthenelse{\boolean{articletitles}}{\emph{{Exploiting CP-asymmetries in rare
  charm decays}}, }{}\href{https://doi.org/10.1103/PhysRevD.101.115006}{Phys.\
  Rev.\ D \textbf{101} (2020) 115006},
  \href{http://arxiv.org/abs/2004.01206}{{\normalfont\ttfamily
  arXiv:2004.01206}}\relax
\mciteBstWouldAddEndPuncttrue
\mciteSetBstMidEndSepPunct{\mcitedefaultmidpunct}
{\mcitedefaultendpunct}{\mcitedefaultseppunct}\relax
\EndOfBibitem
\bibitem{Paul:2011ar}
A.~Paul, I.~I. Bigi, and S.~Recksiegel,
  \ifthenelse{\boolean{articletitles}}{\emph{{On $D\to X_u l^+ l^-$ within the
  standard model and frameworks like the littlest Higgs model with T parity}},
  }{}\href{https://doi.org/10.1103/PhysRevD.83.114006}{Phys.\ Rev.\
  \textbf{D83} (2011) 114006},
  \href{http://arxiv.org/abs/1101.6053}{{\normalfont\ttfamily
  arXiv:1101.6053}}\relax
\mciteBstWouldAddEndPuncttrue
\mciteSetBstMidEndSepPunct{\mcitedefaultmidpunct}
{\mcitedefaultendpunct}{\mcitedefaultseppunct}\relax
\EndOfBibitem
\bibitem{Burdman:2001tf}
G.~Burdman, E.~Golowich, J.~L. Hewett, and S.~Pakvasa,
  \ifthenelse{\boolean{articletitles}}{\emph{{Rare charm decays in the standard
  model and beyond}},
  }{}\href{https://doi.org/10.1103/PhysRevD.66.014009}{Phys.\ Rev.\
  \textbf{D66} (2002) 014009},
  \href{http://arxiv.org/abs/hep-ph/0112235}{{\normalfont\ttfamily
  arXiv:hep-ph/0112235}}\relax
\mciteBstWouldAddEndPuncttrue
\mciteSetBstMidEndSepPunct{\mcitedefaultmidpunct}
{\mcitedefaultendpunct}{\mcitedefaultseppunct}\relax
\EndOfBibitem
\bibitem{Wang:2014dba}
R.-M. Wang {\em et~al.}, \ifthenelse{\boolean{articletitles}}{\emph{{Studying
  the lepton number and lepton flavor violating $D^0\to e^\pm\mu^\mp$ and
  $D^+_{d/s}\to \pi(K)^+e^\pm\mu^\mp$ decays}},
  }{}\href{https://doi.org/10.1142/S0217751X14501693}{Int.\ J.\ Mod.\ Phys.\
  \textbf{A29} (2014) 1450169}\relax
\mciteBstWouldAddEndPuncttrue
\mciteSetBstMidEndSepPunct{\mcitedefaultmidpunct}
{\mcitedefaultendpunct}{\mcitedefaultseppunct}\relax
\EndOfBibitem
\bibitem{Delaunay:2012cz}
C.~Delaunay, J.~F. Kamenik, G.~Perez, and L.~Randall,
  \ifthenelse{\boolean{articletitles}}{\emph{{Charming CP violation and dipole
  operators from RS flavor anarchy}},
  }{}\href{https://doi.org/10.1007/JHEP01(2013)027}{JHEP \textbf{01} (2013)
  027}, \href{http://arxiv.org/abs/1207.0474}{{\normalfont\ttfamily
  arXiv:1207.0474}}\relax
\mciteBstWouldAddEndPuncttrue
\mciteSetBstMidEndSepPunct{\mcitedefaultmidpunct}
{\mcitedefaultendpunct}{\mcitedefaultseppunct}\relax
\EndOfBibitem
\bibitem{Paul:2012ab}
A.~Paul, A.~De~La~Puente, and I.~I. Bigi,
  \ifthenelse{\boolean{articletitles}}{\emph{{Manifestations of warped extra
  dimension in rare charm decays and asymmetries}},
  }{}\href{https://doi.org/10.1103/PhysRevD.90.014035}{Phys.\ Rev.\
  \textbf{D90} (2014) 014035},
  \href{http://arxiv.org/abs/1212.4849}{{\normalfont\ttfamily
  arXiv:1212.4849}}\relax
\mciteBstWouldAddEndPuncttrue
\mciteSetBstMidEndSepPunct{\mcitedefaultmidpunct}
{\mcitedefaultendpunct}{\mcitedefaultseppunct}\relax
\EndOfBibitem
\bibitem{LHCB-PAPER-2012-051}
LHCb collaboration, R.~Aaij {\em et~al.},
  \ifthenelse{\boolean{articletitles}}{\emph{{Search for
  \mbox{\decay{\D^+_{(s)}}{\pip\mumu}} and
  \mbox{\decay{\D^+_{(s)}}{\pim\mup\mup}} decays}},
  }{}\href{https://doi.org/10.1016/j.physletb.2013.06.010}{Phys.\ Lett.\
  \textbf{B724} (2013) 203},
  \href{http://arxiv.org/abs/1304.6365}{{\normalfont\ttfamily
  arXiv:1304.6365}}\relax
\mciteBstWouldAddEndPuncttrue
\mciteSetBstMidEndSepPunct{\mcitedefaultmidpunct}
{\mcitedefaultendpunct}{\mcitedefaultseppunct}\relax
\EndOfBibitem
\bibitem{Lees:2011hb}
BaBar collaboration, J.~P. Lees {\em et~al.},
  \ifthenelse{\boolean{articletitles}}{\emph{{Searches for rare or forbidden
  semileptonic charm decays}},
  }{}\href{https://doi.org/10.1103/PhysRevD.84.072006}{Phys.\ Rev.\
  \textbf{D84} (2011) 072006},
  \href{http://arxiv.org/abs/1107.4465}{{\normalfont\ttfamily
  arXiv:1107.4465}}\relax
\mciteBstWouldAddEndPuncttrue
\mciteSetBstMidEndSepPunct{\mcitedefaultmidpunct}
{\mcitedefaultendpunct}{\mcitedefaultseppunct}\relax
\EndOfBibitem
\bibitem{Rubin:2010cq}
CLEO collaboration, P.~Rubin {\em et~al.},
  \ifthenelse{\boolean{articletitles}}{\emph{{Search for rare and forbidden
  decays of charm and charmed-strange mesons to final states $h^\pm e^\mp
  e^+$}}, }{}\href{https://doi.org/10.1103/PhysRevD.82.092007}{Phys.\ Rev.\
  \textbf{D82} (2010) 092007},
  \href{http://arxiv.org/abs/1009.1606}{{\normalfont\ttfamily
  arXiv:1009.1606}}\relax
\mciteBstWouldAddEndPuncttrue
\mciteSetBstMidEndSepPunct{\mcitedefaultmidpunct}
{\mcitedefaultendpunct}{\mcitedefaultseppunct}\relax
\EndOfBibitem
\bibitem{LHCb-DP-2008-001}
LHCb collaboration, A.~A. Alves~Jr.\ {\em et~al.},
  \ifthenelse{\boolean{articletitles}}{\emph{{The \lhcb detector at the LHC}},
  }{}\href{https://doi.org/10.1088/1748-0221/3/08/S08005}{JINST \textbf{3}
  (2008) S08005}\relax
\mciteBstWouldAddEndPuncttrue
\mciteSetBstMidEndSepPunct{\mcitedefaultmidpunct}
{\mcitedefaultendpunct}{\mcitedefaultseppunct}\relax
\EndOfBibitem
\bibitem{LHCb-DP-2014-002}
LHCb collaboration, R.~Aaij {\em et~al.},
  \ifthenelse{\boolean{articletitles}}{\emph{{LHCb detector performance}},
  }{}\href{https://doi.org/10.1142/S0217751X15300227}{Int.\ J.\ Mod.\ Phys.\
  \textbf{A30} (2015) 1530022},
  \href{http://arxiv.org/abs/1412.6352}{{\normalfont\ttfamily
  arXiv:1412.6352}}\relax
\mciteBstWouldAddEndPuncttrue
\mciteSetBstMidEndSepPunct{\mcitedefaultmidpunct}
{\mcitedefaultendpunct}{\mcitedefaultseppunct}\relax
\EndOfBibitem
\bibitem{Sjostrand:2007gs}
T.~Sj\"{o}strand, S.~Mrenna, and P.~Skands,
  \ifthenelse{\boolean{articletitles}}{\emph{{A brief introduction to PYTHIA
  8.1}}, }{}\href{https://doi.org/10.1016/j.cpc.2008.01.036}{Comput.\ Phys.\
  Commun.\  \textbf{178} (2008) 852},
  \href{http://arxiv.org/abs/0710.3820}{{\normalfont\ttfamily
  arXiv:0710.3820}}\relax
\mciteBstWouldAddEndPuncttrue
\mciteSetBstMidEndSepPunct{\mcitedefaultmidpunct}
{\mcitedefaultendpunct}{\mcitedefaultseppunct}\relax
\EndOfBibitem
\bibitem{Sjostrand:2006za}
T.~Sj\"{o}strand, S.~Mrenna, and P.~Skands,
  \ifthenelse{\boolean{articletitles}}{\emph{{PYTHIA 6.4 physics and manual}},
  }{}\href{https://doi.org/10.1088/1126-6708/2006/05/026}{JHEP \textbf{05}
  (2006) 026}, \href{http://arxiv.org/abs/hep-ph/0603175}{{\normalfont\ttfamily
  arXiv:hep-ph/0603175}}\relax
\mciteBstWouldAddEndPuncttrue
\mciteSetBstMidEndSepPunct{\mcitedefaultmidpunct}
{\mcitedefaultendpunct}{\mcitedefaultseppunct}\relax
\EndOfBibitem
\bibitem{LHCb-PROC-2010-056}
I.~Belyaev {\em et~al.}, \ifthenelse{\boolean{articletitles}}{\emph{{Handling
  of the generation of primary events in Gauss, the LHCb simulation
  framework}}, }{}\href{https://doi.org/10.1088/1742-6596/331/3/032047}{J.\
  Phys.\ Conf.\ Ser.\  \textbf{331} (2011) 032047}\relax
\mciteBstWouldAddEndPuncttrue
\mciteSetBstMidEndSepPunct{\mcitedefaultmidpunct}
{\mcitedefaultendpunct}{\mcitedefaultseppunct}\relax
\EndOfBibitem
\bibitem{Lange:2001uf}
D.~J. Lange, \ifthenelse{\boolean{articletitles}}{\emph{{The EvtGen particle
  decay simulation package}},
  }{}\href{https://doi.org/10.1016/S0168-9002(01)00089-4}{Nucl.\ Instrum.\
  Meth.\  \textbf{A462} (2001) 152}\relax
\mciteBstWouldAddEndPuncttrue
\mciteSetBstMidEndSepPunct{\mcitedefaultmidpunct}
{\mcitedefaultendpunct}{\mcitedefaultseppunct}\relax
\EndOfBibitem
\bibitem{Golonka:2005pn}
P.~Golonka and Z.~Was, \ifthenelse{\boolean{articletitles}}{\emph{{PHOTOS Monte
  Carlo: A precision tool for QED corrections in $Z$ and $W$ decays}},
  }{}\href{https://doi.org/10.1140/epjc/s2005-02396-4}{Eur.\ Phys.\ J.\
  \textbf{C45} (2006) 97},
  \href{http://arxiv.org/abs/hep-ph/0506026}{{\normalfont\ttfamily
  arXiv:hep-ph/0506026}}\relax
\mciteBstWouldAddEndPuncttrue
\mciteSetBstMidEndSepPunct{\mcitedefaultmidpunct}
{\mcitedefaultendpunct}{\mcitedefaultseppunct}\relax
\EndOfBibitem
\bibitem{Allison:2006ve}
Geant4 collaboration, J.~Allison {\em et~al.},
  \ifthenelse{\boolean{articletitles}}{\emph{{Geant4 developments and
  applications}}, }{}\href{https://doi.org/10.1109/TNS.2006.869826}{IEEE
  Trans.\ Nucl.\ Sci.\  \textbf{53} (2006) 270}\relax
\mciteBstWouldAddEndPuncttrue
\mciteSetBstMidEndSepPunct{\mcitedefaultmidpunct}
{\mcitedefaultendpunct}{\mcitedefaultseppunct}\relax
\EndOfBibitem
\bibitem{Agostinelli:2002hh}
Geant4 collaboration, S.~Agostinelli {\em et~al.},
  \ifthenelse{\boolean{articletitles}}{\emph{{Geant4: A simulation toolkit}},
  }{}\href{https://doi.org/10.1016/S0168-9002(03)01368-8}{Nucl.\ Instrum.\
  Meth.\  \textbf{A506} (2003) 250}\relax
\mciteBstWouldAddEndPuncttrue
\mciteSetBstMidEndSepPunct{\mcitedefaultmidpunct}
{\mcitedefaultendpunct}{\mcitedefaultseppunct}\relax
\EndOfBibitem
\bibitem{LHCb-PROC-2011-006}
M.~Clemencic {\em et~al.}, \ifthenelse{\boolean{articletitles}}{\emph{{The
  \lhcb simulation application, Gauss: Design, evolution and experience}},
  }{}\href{https://doi.org/10.1088/1742-6596/331/3/032023}{J.\ Phys.\ Conf.\
  Ser.\  \textbf{331} (2011) 032023}\relax
\mciteBstWouldAddEndPuncttrue
\mciteSetBstMidEndSepPunct{\mcitedefaultmidpunct}
{\mcitedefaultendpunct}{\mcitedefaultseppunct}\relax
\EndOfBibitem
\bibitem{xgboost}
T.~Chen and C.~Guestrin, \ifthenelse{\boolean{articletitles}}{\emph{{XGBoost}:
  A scalable tree boosting system}, }{} in {\em Proceedings of the 22nd ACM
  SIGKDD International Conference on Knowledge Discovery and Data Mining},
  \href{https://doi.org/10.1145/2939672.2939785}{ KDD '16, (New York, NY, USA),
  785--794, ACM, 2016}\relax
\mciteBstWouldAddEndPuncttrue
\mciteSetBstMidEndSepPunct{\mcitedefaultmidpunct}
{\mcitedefaultendpunct}{\mcitedefaultseppunct}\relax
\EndOfBibitem
\bibitem{Cranmer:2000du}
K.~S. Cranmer, \ifthenelse{\boolean{articletitles}}{\emph{{Kernel estimation in
  high-energy physics}},
  }{}\href{https://doi.org/10.1016/S0010-4655(00)00243-5}{Comput.\ Phys.\
  Commun.\  \textbf{136} (2001) 198},
  \href{http://arxiv.org/abs/hep-ex/0011057}{{\normalfont\ttfamily
  arXiv:hep-ex/0011057}}\relax
\mciteBstWouldAddEndPuncttrue
\mciteSetBstMidEndSepPunct{\mcitedefaultmidpunct}
{\mcitedefaultendpunct}{\mcitedefaultseppunct}\relax
\EndOfBibitem
\bibitem{Cowan:2016tnm}
G.~A. Cowan, D.~C. Craik, and M.~D. Needham,
  \ifthenelse{\boolean{articletitles}}{\emph{{RapidSim: an application for the
  fast simulation of heavy-quark hadron decays}},
  }{}\href{https://doi.org/10.1016/j.cpc.2017.01.029}{Comput.\ Phys.\ Commun.\
  \textbf{214} (2017) 239},
  \href{http://arxiv.org/abs/1612.07489}{{\normalfont\ttfamily
  arXiv:1612.07489}}\relax
\mciteBstWouldAddEndPuncttrue
\mciteSetBstMidEndSepPunct{\mcitedefaultmidpunct}
{\mcitedefaultendpunct}{\mcitedefaultseppunct}\relax
\EndOfBibitem
\bibitem{rogozhnikov:2016bdp}
A.~Rogozhnikov, \ifthenelse{\boolean{articletitles}}{\emph{{Reweighting with
  boosted decision trees}},
  }{}\href{https://doi.org/10.1088/1742-6596/762/1/012036}{J.\ Phys.\ Conf.\
  Ser.\  \textbf{762} (2016) 012036},
  \href{http://arxiv.org/abs/1608.05806}{{\normalfont\ttfamily
  arXiv:1608.05806}}\relax
\mciteBstWouldAddEndPuncttrue
\mciteSetBstMidEndSepPunct{\mcitedefaultmidpunct}
{\mcitedefaultendpunct}{\mcitedefaultseppunct}\relax
\EndOfBibitem
\bibitem{Pivk:2004ty}
M.~Pivk and F.~R. Le~Diberder,
  \ifthenelse{\boolean{articletitles}}{\emph{{sPlot: A statistical tool to
  unfold data distributions}},
  }{}\href{https://doi.org/10.1016/j.nima.2005.08.106}{Nucl.\ Instrum.\ Meth.\
  \textbf{A555} (2005) 356},
  \href{http://arxiv.org/abs/physics/0402083}{{\normalfont\ttfamily
  arXiv:physics/0402083}}\relax
\mciteBstWouldAddEndPuncttrue
\mciteSetBstMidEndSepPunct{\mcitedefaultmidpunct}
{\mcitedefaultendpunct}{\mcitedefaultseppunct}\relax
\EndOfBibitem
\bibitem{meerkat}
A.~Poluektov, \ifthenelse{\boolean{articletitles}}{\emph{{Kernel density
  estimation of a multidimensional efficiency profile}},
  }{}\href{https://doi.org/10.1088/1748-0221/10/02/P02011}{JINST \textbf{10}
  (2015) P02011}, \href{http://arxiv.org/abs/1411.5528}{{\normalfont\ttfamily
  arXiv:1411.5528}}\relax
\mciteBstWouldAddEndPuncttrue
\mciteSetBstMidEndSepPunct{\mcitedefaultmidpunct}
{\mcitedefaultendpunct}{\mcitedefaultseppunct}\relax
\EndOfBibitem
\bibitem{LHCb-DP-2018-001}
R.~Aaij {\em et~al.}, \ifthenelse{\boolean{articletitles}}{\emph{{Selection and
  processing of calibration samples to measure the particle identification
  performance of the LHCb experiment in Run 2}},
  }{}\href{https://doi.org/10.1140/epjti/s40485-019-0050-z}{Eur.\ Phys.\ J.\
  Tech.\ Instr.\  \textbf{6} (2018) 1},
  \href{http://arxiv.org/abs/1803.00824}{{\normalfont\ttfamily
  arXiv:1803.00824}}\relax
\mciteBstWouldAddEndPuncttrue
\mciteSetBstMidEndSepPunct{\mcitedefaultmidpunct}
{\mcitedefaultendpunct}{\mcitedefaultseppunct}\relax
\EndOfBibitem
\bibitem{LHCb-DP-2013-002}
LHCb collaboration, R.~Aaij {\em et~al.},
  \ifthenelse{\boolean{articletitles}}{\emph{{Measurement of the track
  reconstruction efficiency at LHCb}},
  }{}\href{https://doi.org/10.1088/1748-0221/10/02/P02007}{JINST \textbf{10}
  (2015) P02007}, \href{http://arxiv.org/abs/1408.1251}{{\normalfont\ttfamily
  arXiv:1408.1251}}\relax
\mciteBstWouldAddEndPuncttrue
\mciteSetBstMidEndSepPunct{\mcitedefaultmidpunct}
{\mcitedefaultendpunct}{\mcitedefaultseppunct}\relax
\EndOfBibitem
\bibitem{CLs}
A.~L. Read, \ifthenelse{\boolean{articletitles}}{\emph{{Presentation of search
  results: The CL$_{\rm s}$ technique}},
  }{}\href{https://doi.org/10.1088/0954-3899/28/10/313}{J.\ Phys.\
  \textbf{G28} (2002) 2693}\relax
\mciteBstWouldAddEndPuncttrue
\mciteSetBstMidEndSepPunct{\mcitedefaultmidpunct}
{\mcitedefaultendpunct}{\mcitedefaultseppunct}\relax
\EndOfBibitem
\bibitem{Junk:1999kv}
T.~Junk, \ifthenelse{\boolean{articletitles}}{\emph{{Confidence level
  computation for combining searches with small statistics}},
  }{}\href{https://doi.org/10.1016/S0168-9002(99)00498-2}{Nucl.\ Instrum.\
  Meth.\  \textbf{A434} (1999) 435},
  \href{http://arxiv.org/abs/hep-ex/9902006}{{\normalfont\ttfamily
  arXiv:hep-ex/9902006}}\relax
\mciteBstWouldAddEndPuncttrue
\mciteSetBstMidEndSepPunct{\mcitedefaultmidpunct}
{\mcitedefaultendpunct}{\mcitedefaultseppunct}\relax
\EndOfBibitem
\end{mcitethebibliography}

\newpage
\centerline
{\large\bf LHCb collaboration}
\begin
{flushleft}
\small
R.~Aaij$^{31}$,
C.~Abell{\'a}n~Beteta$^{49}$,
T.~Ackernley$^{59}$,
B.~Adeva$^{45}$,
M.~Adinolfi$^{53}$,
H.~Afsharnia$^{9}$,
C.A.~Aidala$^{83}$,
S.~Aiola$^{25}$,
Z.~Ajaltouni$^{9}$,
S.~Akar$^{64}$
% ...

{\footnotesize \it

% ...
$ ^{9}$Universit{\'e} Clermont Auvergne, CNRS/IN2P3, LPC, Clermont-Ferrand, France\\
% ...
$ ^{25}$INFN Sezione di Milano, Milano, Italy\\
% ...
$ ^{31}$Nikhef National Institute for Subatomic Physics, Amsterdam, Netherlands\\
% ...
$ ^{45}$Instituto Galego de F{\'\i}sica de Altas Enerx{\'\i}as (IGFAE), Universidade de Santiago de Compostela, Santiago de Compostela, Spain\\
% ...
$ ^{49}$Physik-Institut, Universit{\"a}t Z{\"u}rich, Z{\"u}rich, Switzerland\\
% ...
$ ^{53}$H.H. Wills Physics Laboratory, University of Bristol, Bristol, United Kingdom\\
% ...
$ ^{59}$Oliver Lodge Laboratory, University of Liverpool, Liverpool, United Kingdom\\
% ...
$ ^{64}$University of Cincinnati, Cincinnati, OH, United States\\
% ...
$ ^{67}$Syracuse University, Syracuse, NY, United States\\
% ...
$ ^{83}$University of Michigan, Ann Arbor, United States, associated to $^{67}$\\
\bigskip
% ...
\medskip
}
\end{flushleft}

\end{document}